\documentclass[10pt,twocolumn,amsmath,amssymb,aps,floatfix,preprintnumbers,nofootinbib,nobibnotes,bibnotes]{revtex4-2}
\usepackage{graphicx}
\usepackage{dcolumn}
\usepackage{bm}
\usepackage{hyperref}
\usepackage[mathlines]{lineno}
\usepackage{epsfig}  
\usepackage{color}
\usepackage{slashed}
\usepackage{float}
\usepackage[table]{xcolor}
\usepackage{slashed}
\usepackage{commath}
\usepackage{multirow}

\large

\begin{document}

\title{Imprints of flavor anomalies on neutrino oscillations through dark matter halo}

\author{Ashutosh Kumar Alok}
\email{akalok@iitj.ac.in}
\affiliation{Indian Institute of Technology Jodhpur, Jodhpur 342037, India}

\author{Neetu Raj Singh Chundawat}
\email{chundawat.1@iitj.ac.in}
\affiliation{Indian Institute of Technology Jodhpur, Jodhpur 342037, India}

\author{Arindam Mandal}
\email{mandal.3@iitj.ac.in}
\affiliation{Indian Institute of Technology Jodhpur, Jodhpur 342037, India}

\begin{abstract}
In this work we study the impact of  new physics, stimulated by flavor anomalies, on neutrino oscillations through dense dark matter halo. Inspired by a model where  a Majorana dark matter fermion and two new scalar fields contribute to $b \to s \mu^+ \mu^-$ transition at the one loop level, we study the impact of neutrino-dark matter interaction on the oscillation patterns of ultra-high energy cosmic neutrinos  passing through this muonphilic halo located near the center of Milky Way. We find that due to this interaction, the flavor ratios of neutrinos reaching  earth would be different from that of vacuum oscillations. We also consider   a $Z'$ model driven by $L_{\mu}-L_{\tau}$ symmetry and containing a vector-like fermion as a dark matter candidate. It was previously shown that for such a  model,  the three flavors of neutrinos decouple from each other.  This will render a flavor ratio similar to that of vacuum oscillations. Therefore, the interaction of neutrinos with dense dark matter halo can serve as an important tool to discriminate between flavor models with a {\it dark connection}. 
\end{abstract}
 
\maketitle 

\newpage

\section{Introduction}

The Standard Model (SM) of electroweak interaction can account for a large number of measurements performed till date. Still it cannot be considered as the quintessential theory of fundamental interactions of nature. This is due to the fact that the SM fails to elucidate  several observations which lie at the core of our peregrination to understand the origin and evolution of the universe. This includes matter antimatter asymmetry, existence of dark matter (DM) and dark energy.  The expectation for understanding these observations now hinges upon physics beyond SM.

In recent years, evidence for new physics has started emerging from measurements of several observables in flavor physics. Many of these observables are related to decays induced by the quark level transition $b \to s \ell \ell$ ($\ell=e,\mu$). The most enticing discrepancies are in the measurements of lepton flavor universality (LFU) testing ratios $R_K \equiv \Gamma(B^+ \to K^+ \mu^+ \mu^-)/\Gamma(B^+ \to K^+ e^+ e^-)$ \cite{LHCb:2021trn} and $R_{K^*} \equiv \Gamma(B^0 \to K^{*0} \mu^+ \mu^-)/\Gamma(B^0 \to K^{*0} e^+ e^-)$  \cite{rkstar} which disagree with the SM prediction of $\approx 1$ \cite{Hiller:2003js,Bordone:2016gaq,Isidori:2020acz,Straub:2018kue}  at the level of 3.1 and 2.5$\sigma$, respectively. 

 A 	series of measurements from different experiments at the LHC along with the Belle experiment  have reported deviations from their SM predictions of the angular observable  $P'_5$ in $B \to K^* \, \mu^+\,\mu^-$ decay at the level of 3$\sigma$ \cite{Kstarlhcb1,Kstarlhcb2, sm-angular}. In addition, the measured value of the branching ratio of $B_s \to \phi\, \mu^+\,\mu^-$ decay shows tension with the SM prediction at the level of 3.5$\sigma$ \cite{bsphilhc2} . These atypical measurements can be analysed in a model-independent framework using the language of effective field theory \cite{Descotes-Genon:2013wba,Altmannshofer:2013foa,Alok:2017jgr,Alok:2019ufo,Altmannshofer:2021qrr,Carvunis:2021jga,Alguero:2021anc,Geng:2021nhg,Hurth:2021nsi,Angelescu:2021lln,Alok:2022pjb,SinghChundawat:2022zdf}. Irrespective of the adopted methodology for the treatment of hadronic uncertainties, the global analysis of data favours new physics scenarios with non-zero value(s) of the muonic Wilson coefficients.
 
Another evidence for new physics emanates from the measurement of the anomalous magnetic moment of the muon, $a_{\mu}\equiv (g_{\mu}-2)/2$. In 2021, the muon $(g-2)$ experiment at Fermilab E989 announced a new measurement which confirms the enduring discrepancy of $a_{\mu}$ with its SM prediction  \cite{Abi:2021gix}.  The measurement from the previous muon $(g-2)$  experiment at BNL, E821, deviated from the SM value at the level of 3.7$\sigma$ \cite{Bennett:2006fi} whereas the E989 Run 1 data is 3.4$\sigma$ away from the SM expectation.  The combined data gives $a_{\mu}=a_{\mu}^{\rm exp} -a_{\mu}^{\rm SM} = (251 \pm 59) \times 10^{-11}$ implying  an inconsistency with the SM at 4.2$\sigma$ level.

These anomalies can be reconciled in the context of several extensions of the SM. There are a surfeit of such new physics models. Many of these models are linked to the dark sector.  These models can be classified into two categories:  portal and loop models \cite{Vicente:2018frk,London:2021lfn}. In portal models, the mediator of relic density is the same as that of $b \to s \ell \ell$ transition whereas in the loop models, the DM particle contributes to $b \to s \ell \ell$ process at the loop level. In this work, we consider two of  these models and study their impact on oscillations of high energy cosmic neutrinos passing through a dense DM halo. We assume that the DM halo is located near the center of Milky Way galaxy.

Ultra high energy (UHE) neutrinos generated in distant cosmic accelerators provide us an opportunity to probe physics beyond the current paradigm. The potential of these neutrinos ranges from probing non-standard interactions up to Planck scale physics, see for e.g., \cite{Gandhi:1998ri,Bhattacharya:2010xj,Hooper:2004xr,Ackermann:2022rqc,Farzan:2018pnk,Farzan:2021gbx,Farzan:2021slf,Pandey:2018wvh,Karmakar:2020yzn,Brdar:2017kbt}. The prominent source of UHE cosmic neutrons are high energy protons in cosmic rays, $10^{10}$ - $10^{11}$ GeV,   which interact with photons to produce neutrinos in the chain, $p \gamma \to \Delta^+ \to \pi^+ n$, $\pi^+ \to \mu^+ \nu_{\mu}  \to e^+ \nu_e \nu_{\mu} \bar{\nu_{\mu}}$. The energy of these neutrinos can lie in the range of $10^4$ - $10^{10}$ GeV depending upon the type of source which can be of galactic and well as extragalactic origin. 

In 2013, the IceCube Collaboration observed  two  $\sim$ 1 PeV neutrinos, with a $p$-value 2.8$\sigma$ beyond the hypothesis that these observed events were  emanated in the atmosphere \cite{IceCube:2013cdw}. Extragalactic cosmic ray accelerators are considered to be the most viable sources for the neutrinos observed by IceCube \cite{Anchordoqui:2013dnh}. The extragalactic  sources such as gamma-ray bursts, active galactic nuclei, starburst galaxies and newly-born pulsars can potentially generate neutrinos in the PeV and higher energy range \cite{Anchordoqui:2013dnh}. Apart from  IceCube, detectors such as ANTARES \cite{ANTARES:2011hfw}, KM3NeT \cite{KM3Net:2016zxf}, Baikal-GVD \cite{Avrorin:2022lyk}, P-ONE \cite{P-ONE:2020ljt,Resconi:2021ezb}, IceCube-Gen2 \cite{IceCube-Gen2:2020qha,IceCube:2014gqr}, Trinity \cite{Brown:2021lef}, RET \cite{RadarEchoTelescope:2021rca}, TAMBO \cite{Romero-Wolf:2020pzh} have the potential to observe UHE neutrino events.

The spatial distribution of DM within  galaxies is unknown. However, in a number of models, the DM density is assumed to be higher near the center of galaxies as compared to the outer regions. In such models, the density profile of the DM is expressed in terms of the radial distance from the galactic center. 
There are few DM density profile models \cite{penacchioni:2020testing} based on the observed galaxy rotation curves. These include  Navarro–Frenk–White (NFW)\cite{Navarro:1995iw,Navarro:1996gj,Sofue:2011kw}, Einasto\cite{Einasto:1965czb}, isotropic \cite{iso} and RAR \cite{Arguelles:2016ihf,arguelles:2018novel} models. Out of these, the RAR model predicts DM density near the center to be much higher as compared to the other models.

 The interaction of neutrinos with the DM halo can modify the oscillation pattern as well as the flavor compositions and hence it would be interesting to see whether one can discriminate between some of these new physics models based on the modulated oscillation patterns and flavor compositions. Here we consider   a model with two new scalar fields and a Majorana dark matter fermion contributing to the quark level transition $b \to s $  at the one loop level \cite{Cerdeno:2019vpd}. We also consider a  model with  $Z'$ driven by $L_{\mu} - L_{\tau}$ symmetry  and  a vector-like fermion which is a DM particle. This model  can accommodate the current measurement  of muon $(g-2)$ \cite{Chao:2020qpe}.

To explore the discriminating capabilities of a DM halo, we assume a UHE (100 TeV to 1 PeV) cosmic neutrino flux source situated near the center of the Milky Way. This is also the scenario that may render the maximum effect since the center of the DM contains the highest density of DM. If the neutrino flux passes through the arms of the galaxy, where the DM density is much smaller compared to the center, the modulation in flavor composition would be too small to be observed. The radius of the DM halo is considered to be $10^{-3}$ pc. However, owing to the small size of this immoderately densed DM halo, it can be referred to as a sub-halo. The effects of neutrino-DM interaction within the sub-halo can be observed on the Earth only if  the flavor ratios of neutrinos reaching Earth are different from that of  vacuum oscillations \footnote{It should be noted that the neutrino propagation can also start from the opposite side of the sub-halo rather than the center.  Further, these roving fluxes of neutrinos may also run into multiple sub-halos before reaching the Earth. This may also invoke additional contributions to the modulation in flavor compositions. However, for simplicity,  we assume that after emerging out of the DM sub-halo, the neutrinos traverse the rest
path through vacuum.}.

The plan of the paper is as follows. In Sec.\ref{formalism}, we discuss the theoretical framework for change in oscillation patterns for neutrino oscillations after passing through a dense DM sub-halo. This is followed by a brief discussion of various new physics models considered in this work in Sec. \ref{fm}. The results are presented in Sec. \ref{results}. The conclusions are provided in Sec. \ref{conc}.

\section{Neutrino Oscillation in muonphilic DM}
\label{formalism}

In this section we provide a model independent analytical description of neutrino oscillation through a dense muonphilic DM sub-halo. The Hamiltonian governing the dynamics of the three flavor neutrino oscillations is
\begin{equation}
          \mathcal{H}_{0}=\frac{1}{2E}\left[V
                        \begin{pmatrix}
                               0 & 0 & 0\\
                               0 & \Delta m{_{21}^{2}} & 0\\
                               0 & 0 & \Delta m{_{31}^{2}}
                       \end{pmatrix}V^{\dagger}\right],
\end{equation}
where $V$ is the PMNS matrix, $\Delta m^{2}_{ji}=m_{j}^{2}-m_{i}^{2}$ and $E$ is the neutrino energy.  For $V$, we have used the convention of PDG \cite{pdg}. 
Normal mass ordering is used throughout this work. The oscillation parameters \cite{Esteban:2018azc} used in our analysis are given in Table~\ref{parameters}.
\begin{table}[h!]
    \centering
    \begin{tabular}{c|c|c|c|c|c}
    \hline
    $\theta_{12}$ & $\theta_{13}$ & $\theta_{23}$ & $\delta$ & $\Delta m_{21}^{2}$ & $\Delta m_{31}^{2}$  \\
    \hline
    \hline
     $33.82^\circ$ & $8.61^\circ$ & $49.7^\circ$ & $217^\circ$ & $7.39\times 10^{-5}$ & $2.451\times 10^{-3}$\\
     \hline
    \end{tabular}
    \caption{The oscillation parameters for Normal Mass Ordering \cite{pdg}. The mass squared differences are in $\rm eV^2$. }
    \label{parameters}
\end{table}

The neutrino oscillation probabilities in vacuum are given by
\begin{eqnarray}
    P_{\alpha\beta} &=& \delta_{\alpha\beta}-4\sum_{j>i}{\rm Re}[V_{\alpha i}V_{\beta i}^{*}V_{\alpha j}^{*}V_{\beta j}]\sin^{2}\Delta_{ji}\nonumber\\
   && \pm 2 \sum_{j>i}{\rm Im}[V_{\alpha i}V_{\beta i}^{*}V_{\alpha j}^{*}V_{\beta j}]\sin{2\Delta_{ji}},
\end{eqnarray}
where $\Delta_{ji}=\frac{\Delta m^{2}_{ji} L }{4E}$. 
The presence of matter gives rise to the following matrix \cite{luo2020:neutrino}
\begin{equation}
          \mathcal{H}_{M}=\frac{1}{2E}
                       \begin{pmatrix}
                               A_{CC} & 0 & 0\\
                               0 & 0 & 0\\
                               0 & 0 & 0
                       \end{pmatrix},
     \end{equation}
     where $A_{CC}=2\, E\, V_{CC}$. Here $V_{CC}$ is the potential of charge current interaction with electrons. 
 While propagating through the DM sub-halo, the neutrino flux encounters an additional effective potential which generates another term in the Hamiltonian. Thus the total Hamiltonian is given by
      \begin{equation}
        \mathcal{H}=\mathcal{H}_{0}+\mathcal{H}_{M}+\mathcal{H}_{\chi}\,.
      \end{equation}
The form of $\mathcal{H}_{\chi}$ depends upon the nature of the DM. For DM charged under $L_{\mu}-L_{\tau}$, $\mathcal{H}_{\chi}$ given as \cite{Chao:2020qpe}
 \begin{equation}
          \mathcal{H_{\chi}}=\frac{1}{2E}
                       \begin{pmatrix}
                               0 & 0 & 0\\
                               0 & A_{\chi} & 0\\
                               0 & 0 & -A_{\chi}
                       \end{pmatrix}\,.
 \end{equation}
Here $A_{\chi}=2\, E\, V_{\chi}$ and $V_{\chi}$ is the dark matter potential similar to $V_{CC}$ in the case of ordinary matter. On the other hand if  DM interacts only with muon-neutrinos \cite{Cerdeno:2019vpd},  the form of  $\mathcal{H}_{\chi}$ changes to

\begin{equation}
          \mathcal{H_{\chi}}=\frac{1}{2E}
                       \begin{pmatrix}
                               0 & 0 & 0\\
                               0 & A_{\chi} & 0\\
                               0 & 0 & 0
                       \end{pmatrix}\,.
 \end{equation} 
 The Hamiltonian in DM, thus depends on the nature of the DM particle. Throughout this paper, muonphilic is referred to the DM which interacts only with muon-neutrinos (unlike the case of  $L_{\mu}-L_{\tau}$ ).

We consider  propagation of neutrino flux through a DM sub-halo where the DM potential dominates over the ordinary matter potential  i.e. $A_{CC}<<A_{\chi}$.
The validation of this approximation is provided in the next section, where we introduce the DM models. Thus the Hamiltonian reduces to $\mathcal{H}=\mathcal{H}_{0}+\mathcal{H}_{\chi}$. This Hamiltonian, in general,  can be written as 
    \begin{equation}
          \mathcal{H}=\frac{1}{2E}\left[\tilde{V}
                        \begin{pmatrix}
                               \tilde{m}_{1}^{2} & 0 & 0\\
                               0 & \tilde{m}_{2}^{2} & 0\\
                               0 & 0 &  \tilde{m}_{3}^{2}
                       \end{pmatrix}\tilde{V}^{\dagger}
                       \right]\,,
     \label{ham-dm}                  
     \end{equation}
where $\tilde{V}$ is the PMNS matrix and $\tilde{m}_{i}$'s are the eigenvalues of the Hamiltonian in the presence of DM \eqref{ham-dm}. This $\tilde{V}$  and $\tilde{m}_{i}$'s play the role of $V$ and $m_{i}$'s, respectively in the DM environment. Therefore  the survival and oscillation probabilities in the DM sub-halo can be expressed in terms of $\tilde{V}$  and $\tilde{m}_{i}$'s.

Note that eq.~\eqref{ham-dm} represents a generic Hamiltonian. The nature of the DM will be reflected in the $A_{\chi}$-dependence of  $\tilde{V}$  and $\tilde{m}_{i}$. A generic procedure to obtain PMNS matrix elements and oscillation probabilities in DM, starting from the Hamiltonian is given in next two subsections.

\subsection{Mixing matrix elements in DM}
As discussed above, we need the $\tilde{V}$  and $\tilde{m}_{i}$ to obtain the probabilities. One can get the values of PMNS matrix, in the presence of DM, from the eigenvector-eigenvalue identity\cite{Denton:2019pka}
\begin{equation}
 \abs{\tilde{V}_{\alpha_i}}^{2} = \frac{(\lambda_{i}-\xi_{\alpha})(\lambda_{i}-\zeta_{\alpha})}{(\lambda_{i}-\lambda_{j})(\lambda_{i}-\lambda_{k})}, 
\end{equation}
where $\alpha$ and $(i,\, j,\, k)$ denote the flavor and mass basis indices, respectively. $\lambda_i/2E$ and $(\xi_{\alpha}/2E,\, \zeta_{\alpha}/2E)$ are the eigenvalues of $\mathcal{H}$ and $\mathcal{H_{\alpha}}$,  respectively. Here $\mathcal{H_{\alpha}}$ is a $2\times2$ submatrix constructed from $\mathcal{H}$ by removing the $\alpha$-th row and  $\alpha$-th column and can be represented as, 
\begin{equation}
    \mathcal{H_{\alpha}}=\begin{pmatrix}
                            \mathcal{H_{\beta\beta}} & \mathcal{H_{\beta\gamma}}\\ 
                            \mathcal{H_{\gamma\beta}} & \mathcal{H_{\gamma\gamma}}
                         \end{pmatrix}\,.
\end{equation}

The mixing angles in the DM can now be obtained as
\begin{equation}
    \tilde{s}_{12} = \frac{\abs{\tilde{V}_{e2}}}{\sqrt{1-\abs{\tilde{V}_{e3}}^{2}}},\quad
    \tilde{s}_{13}=\abs{\tilde{V}_{e3}}, \quad
     \tilde{s}_{23}=\frac{\abs{\tilde{V}_{\mu2}}}{\sqrt{1-\abs{\tilde{V}_{e3}}^{2}}}\,.
\end{equation}
The phase $\delta$ will also change in DM. The modified $\tilde{\delta}$ can be obtained with the help of the Jarlskog invariant
\begin{equation}
    \mathcal{J}={\rm Im}(V_{e1}V_{\mu1}^{*}V_{e2}^{*}V_{\mu2})=s_{23}c_{23}s_{13}c_{13}^{2}s_{12}c_{12}\sin{\delta}\,,
\end{equation}
which follows the identity
\begin{equation}
    \tilde{\mathcal{J}}\Delta{\tilde{m}_{21}^{2}}\Delta{\tilde{m}_{31}^{2}\Delta{\tilde{m}_{32}^{2}}}=\mathcal{J}\Delta m_{21}^{2}\Delta m_{31}^{2}\Delta m_{32}^{2}.
\end{equation}
Here $\tilde{\mathcal{J}}$ is the Jarlskog in the DM. From these relations one can extract the $\tilde{\delta}$, which is required to compute the oscillation probabilities in DM.

\subsection{Oscillation probabilities in DM}

In the form of the components obtained in previous subsection, the neutrino oscillation probabilities in DM can be written as
\begin{eqnarray}
    \tilde{P}_{\alpha\beta} &=& \tilde{\delta}_{\alpha\beta}-4\sum_{j>i}Re[\tilde{V}_{\alpha i}\tilde{V}_{\beta i}^{*}\tilde{V}_{\alpha j}^{*}\tilde{V}_{\beta j}]\sin^{2}{\tilde{\Delta}_{ji}}\nonumber\\
  &&  \pm 2 \sum_{j>i}{\rm Im}[\tilde{V}_{\alpha i}\tilde{V}_{\beta i}^{*}\tilde{V}_{\alpha j}^{*}\tilde{V}_{\beta j}]\sin{2\tilde{\Delta}_{ji}} ,
\end{eqnarray}
where, similar to the vacuum oscillation, $\tilde{\Delta}_{ji}=\frac{\Delta \tilde{m}^{2}_{ji} L }{4E}$. 
As one can see, the quantities characterizing the neutrino oscillations phenomena changes in the presence of DM and from the previous two subsections, it is  clear that these changes depend upon the type of DM. In the next section, we introduce the DM  models under consideration.

\section{Flavor Models with dark connection}
\label{fm}
In this section, we consider a model with a Majorana dark matter particle along with two scalar fields. The $b \to s \mu^+ \mu^-$ transition is generated at the loop level in this model. 
 We then consider a  $Z'$ model driven by $L_{\mu}-L_{\tau}$ symmetry which can accommodate the current measurement of muon $(g-2)$. This model also includes a vector-like fermion which is a DM particle. 
 In the following, we provide a brief introduction to  these models.

\subsection{A model with scalar fields and a Majorana dark matter fermion for $b \to s \mu^+ \mu^-$}
\label{phi}
We  consider a model with  a Majorana dark matter particle, $\chi$, and two  scalar fields, $\phi_q$ and $\phi_l$, where the scalar field $\phi_q$ has a colour charge \cite{Cerdeno:2019vpd}. This is a modified version of model considered in \cite{Cline:2017qqu}, where the fermionic dark matter field was accompanied by a scalar and a  coloured fermion. 
 In this model, the interaction between the new particles and the SM are described by the following  Lagrangian
\begin{equation}
  \mathcal{L} = \lambda_{Q_i}\bar{Q}_{i}\phi_{q}P_{R}\chi + \lambda_{L_i}\bar{L}_{i}\phi_{l}P_{R}\chi + \rm{h.c.} \, .
  \label{model-lag}
\end{equation}
Here $Q_i$ and $L_i$ are the SM left-handed quark and lepton doublets of each generation, respectively whereas $\lambda_{Q_i}$ and $\lambda_{L_i}$ are the corresponding new physics couplings.  The stability of the DM particle is achieved by imposing  a ${\cal Z}_2$ parity and assuming $m_{\phi_{q,l}} > m_{\chi}$. Under ${\cal Z}_2$ symmetry, the SM fields are invariant. The quantum numbers of the new fields along with the charges under ${\cal Z}_2$ symmetry are presented in Table~\ref{tab:1}. In \cite{Cerdeno:2019vpd}, it was shown that this model can accommodate the current measurements in $b \to s \mu^+ \mu^-$  sector along with reproducing the observed dark matter relic density provided the new physics  couplings are complex.
\begin{table}[h]
\begin{center}
\begin{tabular}{|c|c|c|c||c|}
\hline  
& $SU(3)$ & $SU(2)_{L}$ & $U(1)_{Y}$ & ${\cal Z}_2$ \\
\hline
\hline $\phi_{q}$ & $3$ & $2$ & $1/6$ & $-1$ \\
\hline $\phi_{l}$ & $1$ & $2$ & $-1/2$ & $-1$ \\
\hline $\chi$ & $1$ & $1$ & $0$ & $-1$ \\
\hline
\end{tabular}
\caption{Quantum numbers of the new fields along with the charges under ${\cal Z}_2$ symmetry.}
\label{tab:1}
\end{center}
\end{table}

The fact that the new physics impressions are mainly emerging out in the interactions where muons are involved, in this model, the coupling of the new particles with only muons is considered. This new particles in this model contributes to $b \to s \mu \mu$ processes at one loop level. Due to SU(2) symmetry, muon neutrinos can also interact with the DM. The interaction following the Lagrangian \eqref{model-lag} engenders the following effective potential
\begin{equation}
    V_{\chi}=\left(\frac{\lambda_{\mu}}{4m_{\phi_{\mu}}}\right)^{2}n_{\chi}\,,
    \label{dmpot} 
\end{equation}
where $\lambda_{\mu}=\sqrt{4\pi}$ and $m_{\phi_{\mu}}$ is the mass of the scalar mediator.

\subsection{$Z'$ for $(g-2)_{\mu}$}
\label{z}
The experimental value of anomalous magnetic moment of muon deviates from its SM prediction at the level of   $4.2\sigma$ \cite{  Abi:2021gix,Bennett:2006fi}. This difference can be attributed to new physics contributions coming from beyond SM  particle(s). In\cite{Chao:2020qpe}, a $U(1)_{L_{\mu}-L_{\tau}}$ model was considered to explain the muon $(g-2)_{\mu}$ anomaly. This model  also included a DM particle. In this model, the Lagrangian additional to the SM Lagrangian is  given as
\begin{equation}
   \mathcal{L} = \overline{\chi}i\slashed{D}\chi + (D_{\mu}\Phi)^{\dagger}(D^{\mu}\Phi) - m_{\chi}\overline{\chi}\chi + {\mu}^{2}{\Phi}^{\dagger}\Phi - \lambda({\Phi}^{\dagger}\Phi)^{2}\,,
\end{equation} 
where $D_{\mu} = \partial_{\mu} - ig_{X}Z^{'}_{\mu}$ is the covariant derivative corresponding to the new gauge boson $Z'$, with $g_{X}$ being the new gauge coupling. $\chi$ is a vector-like fermion and it is considered to be a DM candidate whereas $\Phi$ is a complex scalar singlet, having a non-zero vacuum expectation value which gives rise to the mass of $Z'$ when the spontaneous symmetry of the $U(1)_{L_{\mu}-L_{\tau}}$ gauge symmetry is broken. 
SM particles charged under this model can interact to the DM particle $\chi$ via the new mediator $Z'$. Thus $\mu$- and $\tau$-neutrinos can interact with the DM. This interaction can induce $A_{\chi}$, introduced  in the previous section.
The effective potential generated due to the DM interaction with $\nu_{\mu}$ and $\nu_{\tau}$ is given by
\begin{equation}
    V_{\chi} = \pm \left(\frac{g_{X}}{m_{Z'}}\right)^{2}n_{\chi}\,,
    \label{dmlpot}
\end{equation}
where the positive and negative sign corresponds to $\nu_{\mu}$ and $\nu_{\tau}$, respectively (due to their charge under the $U(1)_{L_{\mu}-L_{\tau}}$ model). Here $n_{\chi}$ is the number density of DM.

We work under the assumption that the DM sub-halo consists only of  the DM particles and neglect the presence of anti-DM. A DM asymmetry of this type can be accompanied with the baryo- and leptogenesis, generating asymmetry in both the SM and dark sector, see for e.g., \cite{Zurek:2013wia}. However, a detailed analysis of the generation of this type of asymmetry is beyond the scope of this work.

\begin{figure*}[h!]
\begin{center}
    \includegraphics[width=0.3\textwidth]{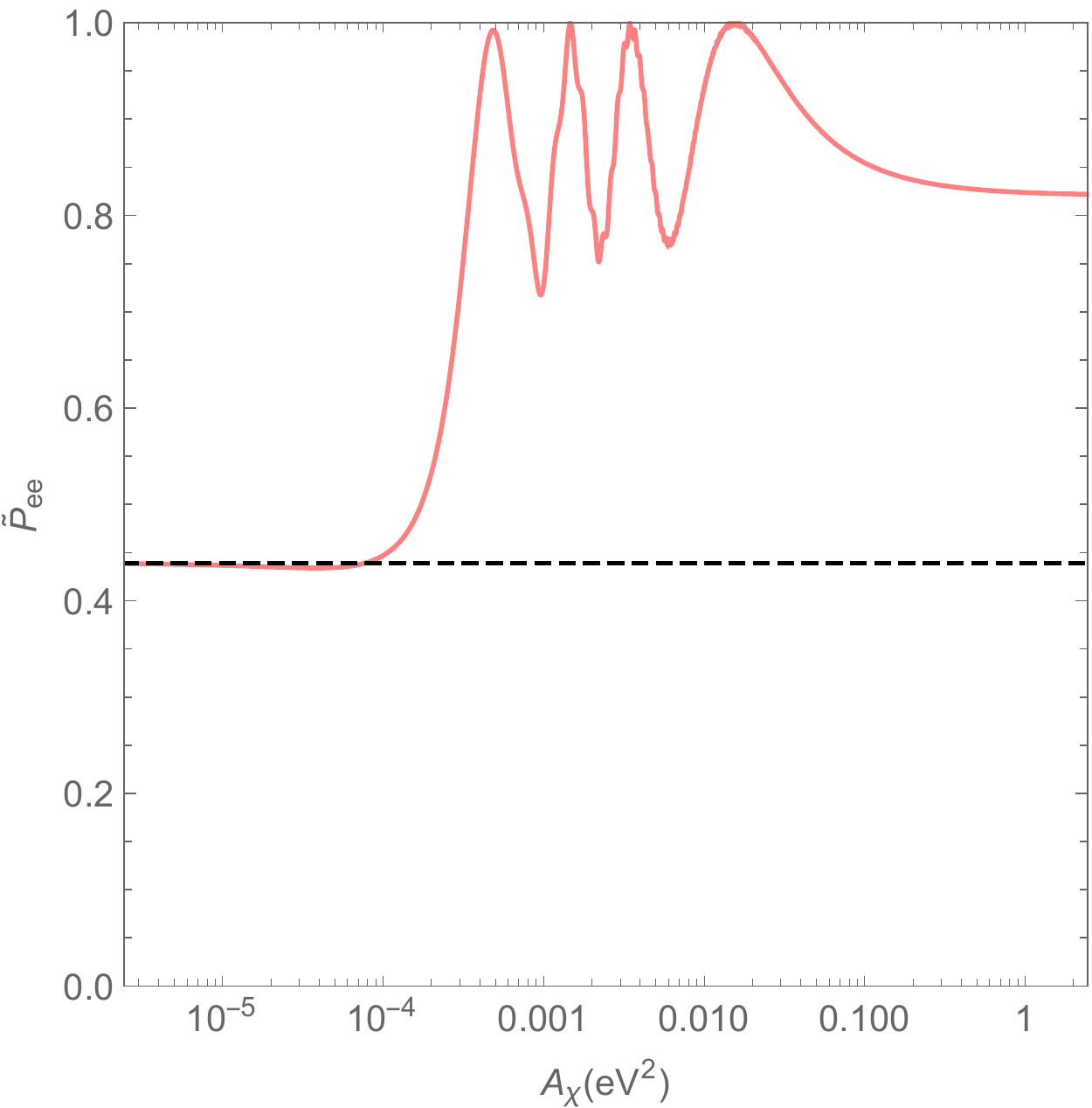} 
    \includegraphics[width=0.3\textwidth]{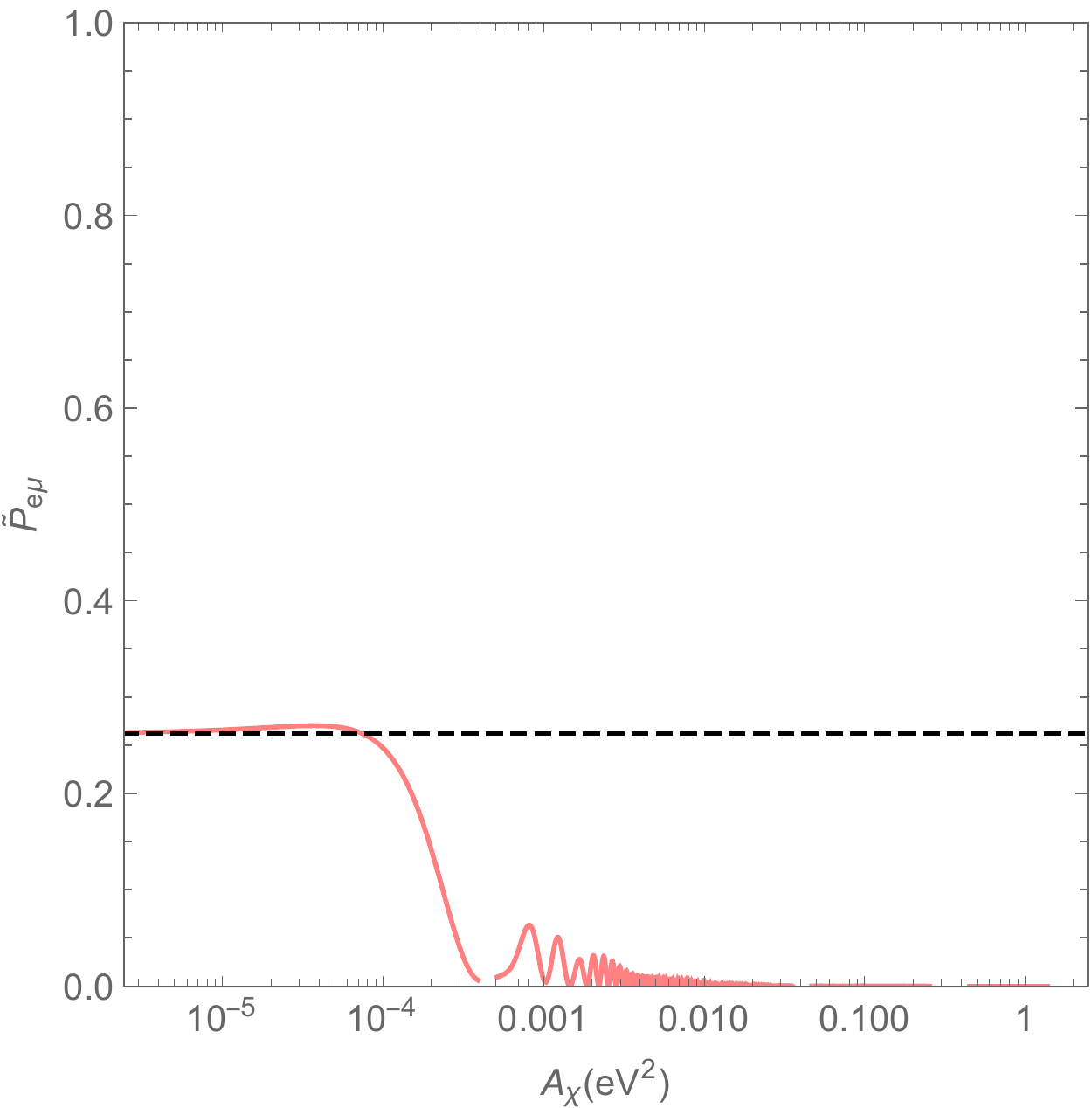}
     \includegraphics[width=0.3\textwidth]{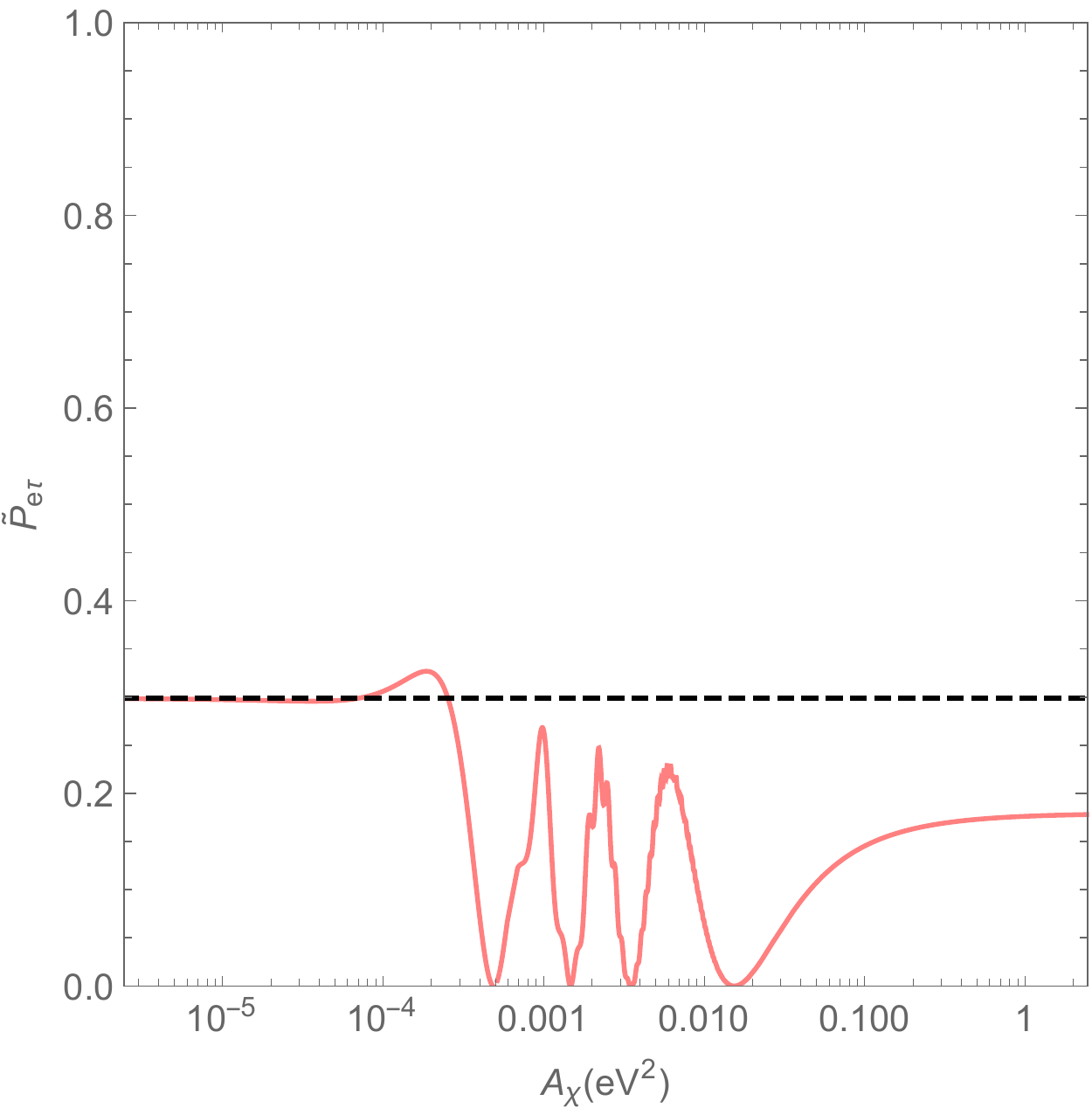} \\ 
    \includegraphics[width=0.3\textwidth]{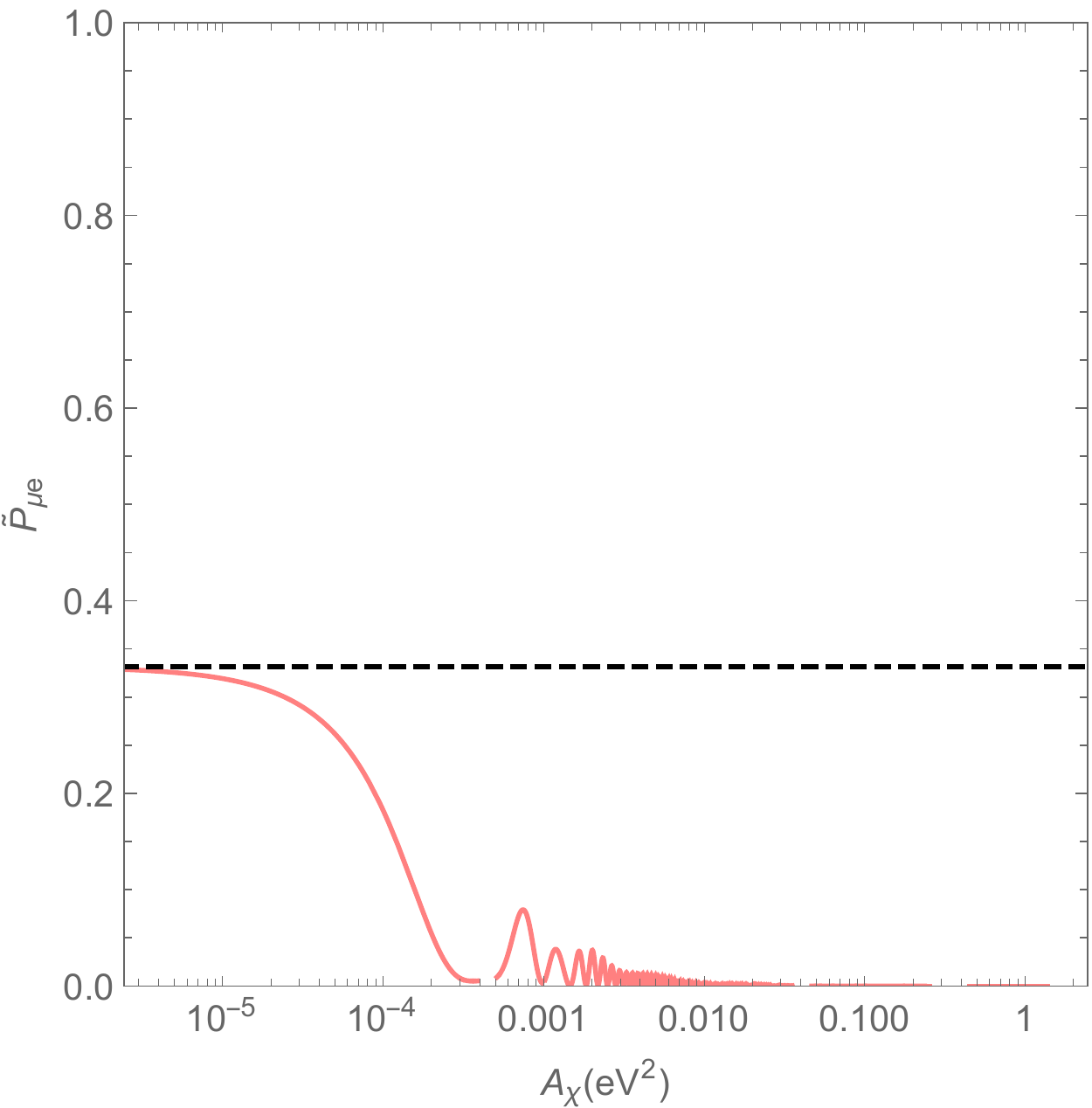} 
    \includegraphics[width=0.3\textwidth]{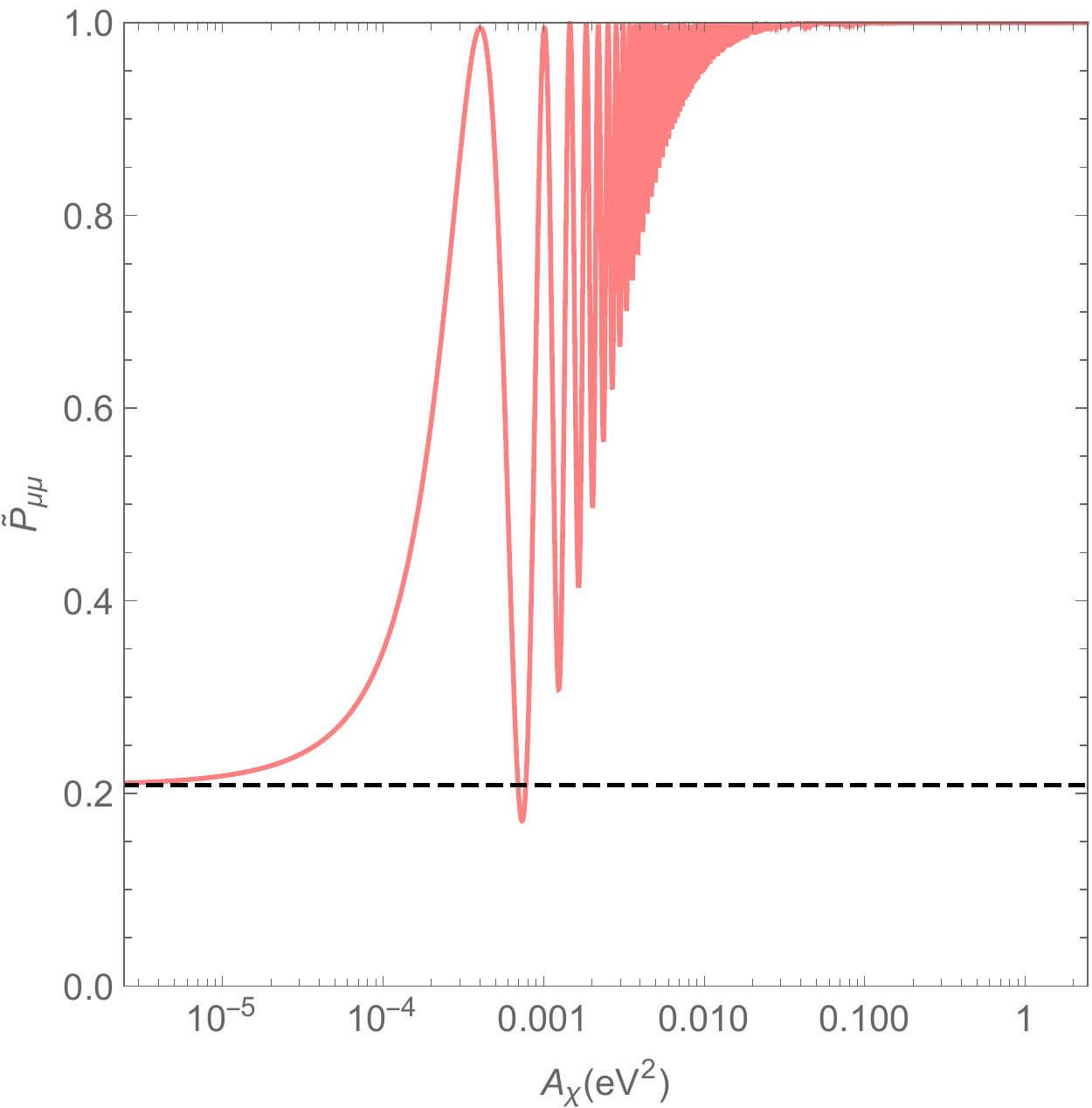}
     \includegraphics[width=0.3\textwidth]{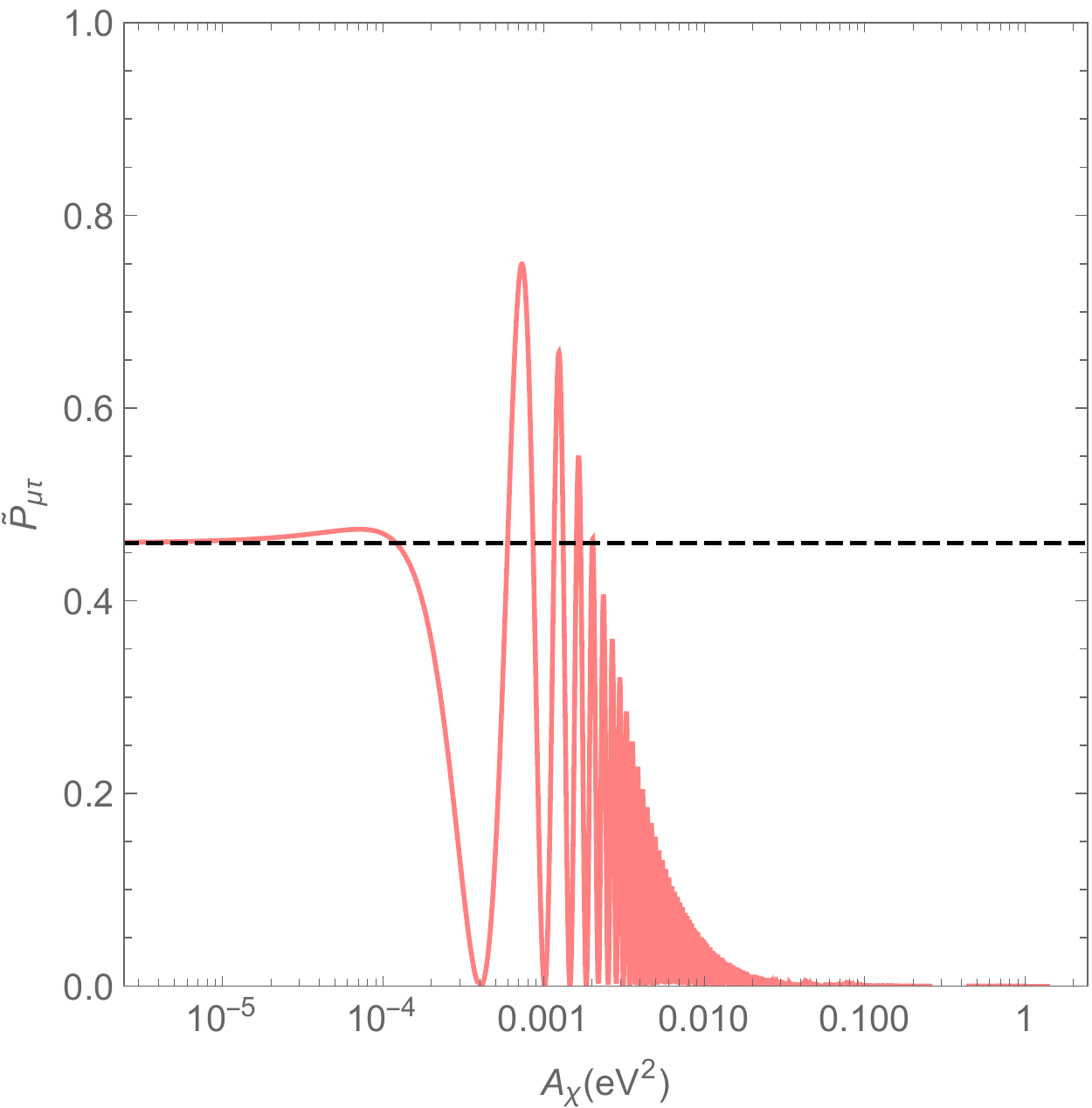}\
     \includegraphics[width=0.3\textwidth]{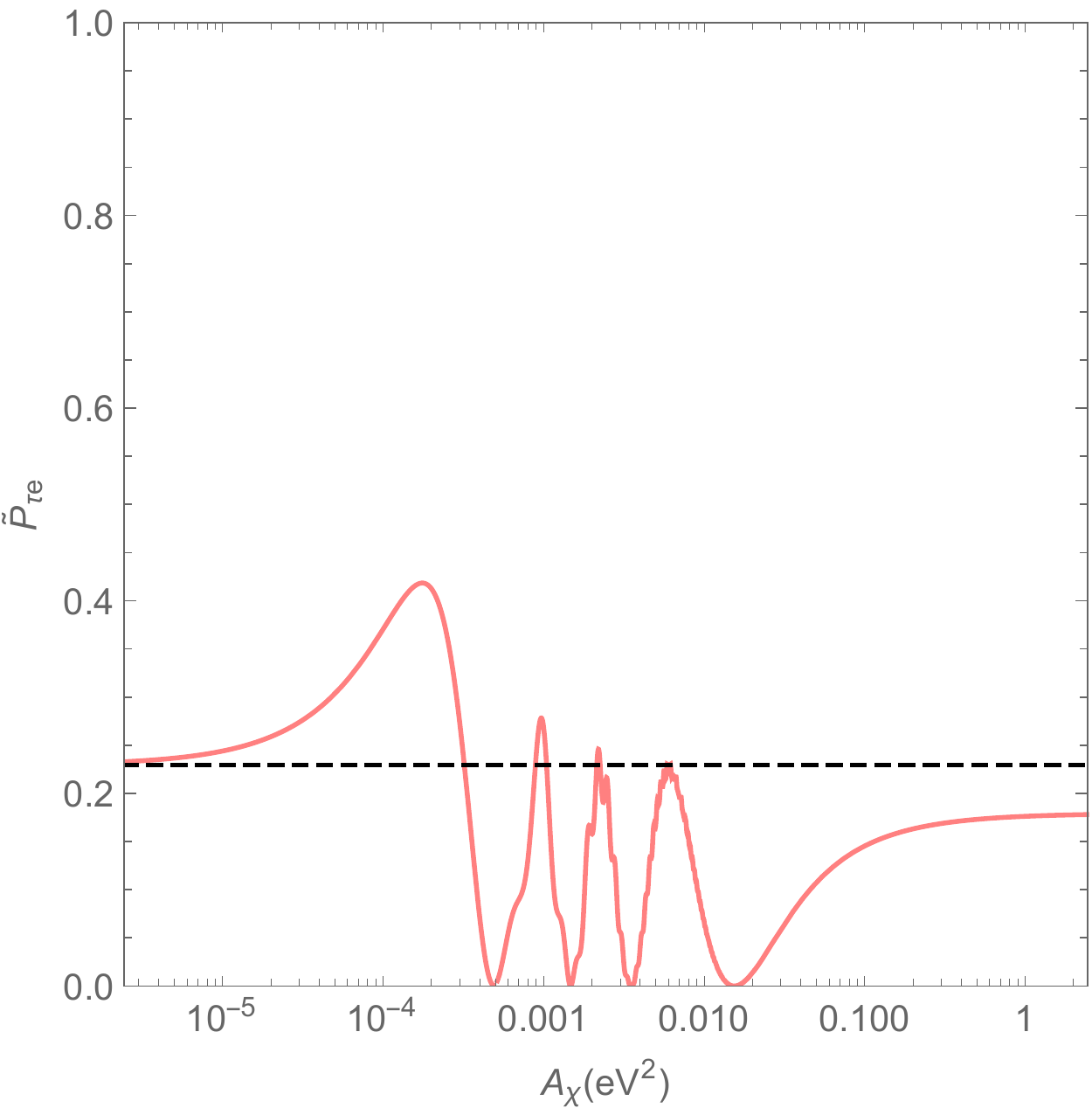} 
    \includegraphics[width=0.3\textwidth]{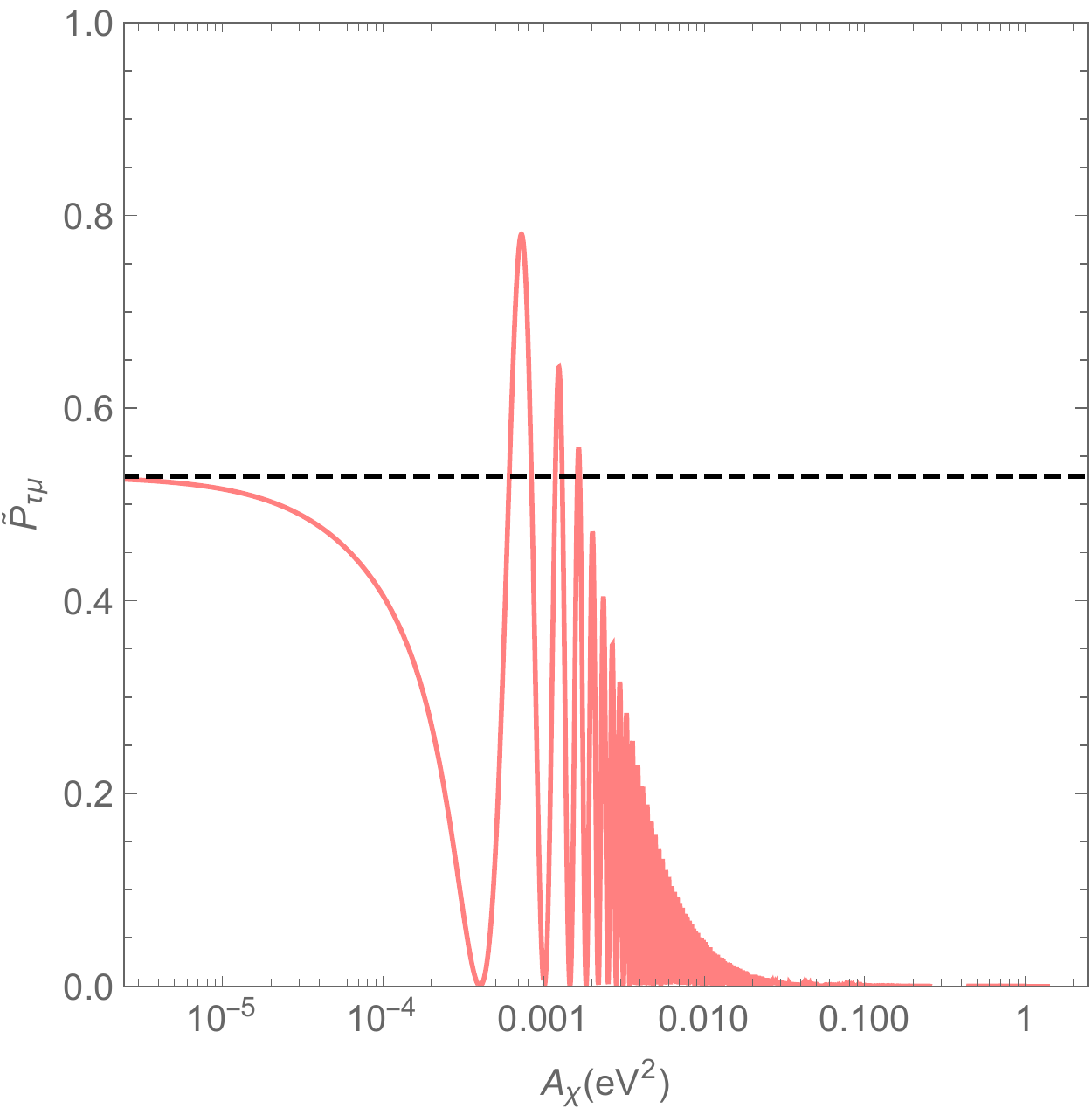}
     \includegraphics[width=0.3\textwidth]{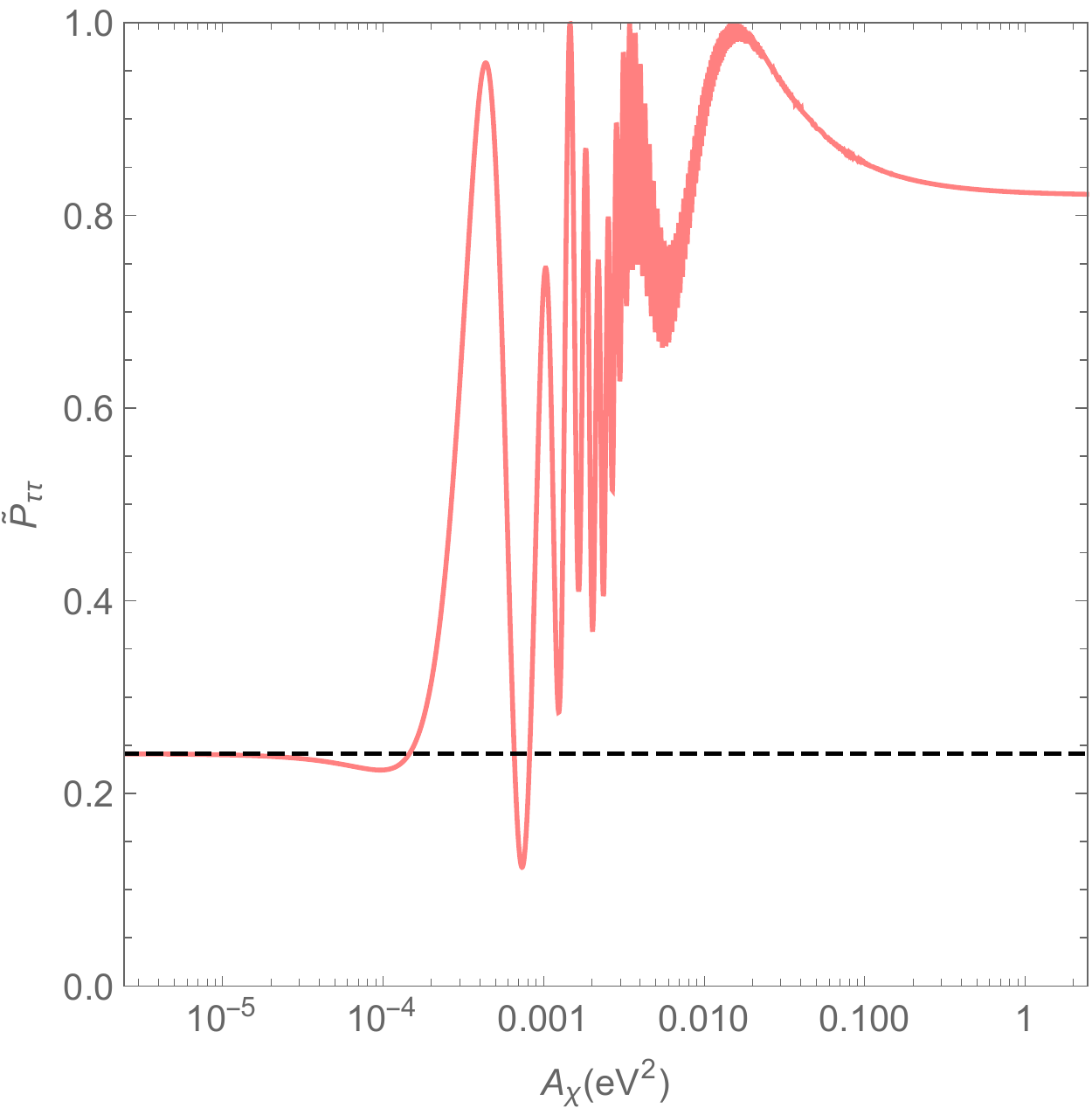}\\
   \end{center}
   \caption{Neutrino oscillation probabilities as a function of $A_{\chi} \, \rm (eV^2)$ for muonphilic DM. Black dashed line represents the probability values in vacuum. For low values of $A_{\chi}$, the oscillation probabilities are close to the vacuum oscillation probabilities. In the mid-$A_{\chi}$, probabilities start to deviate from the vacuum. In the high-$A_{\chi}$ region, $\nu_{\mu}$ decouples from the other flavors. However $\nu_{e}-\nu_{\tau}$ is oscillation still possible.}
   \label{nprob-1}
   \end{figure*}

\begin{figure*}[hbt]
\begin{center}
\includegraphics[width=0.3\textwidth]{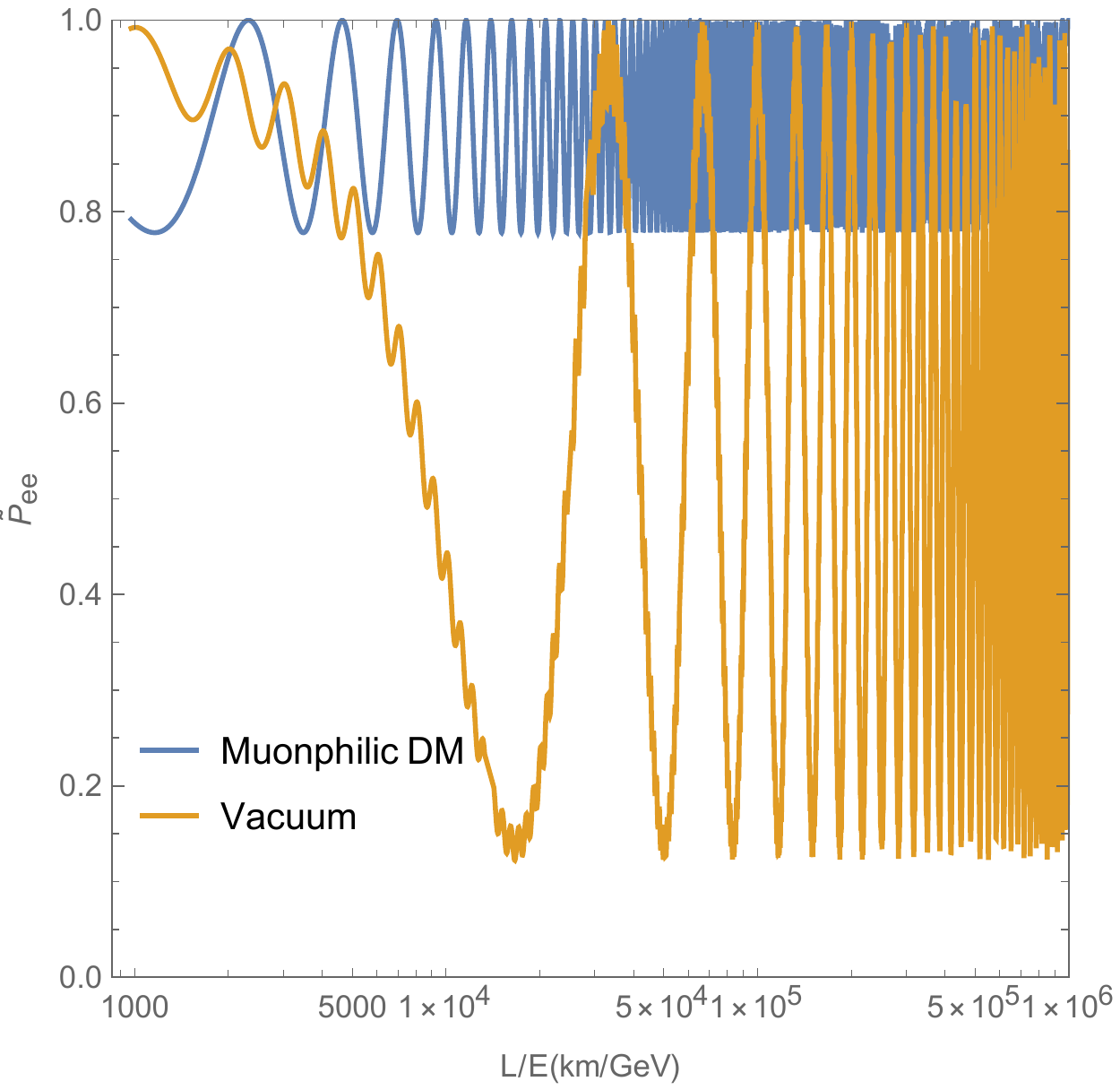} 
\includegraphics[width=0.3\textwidth]{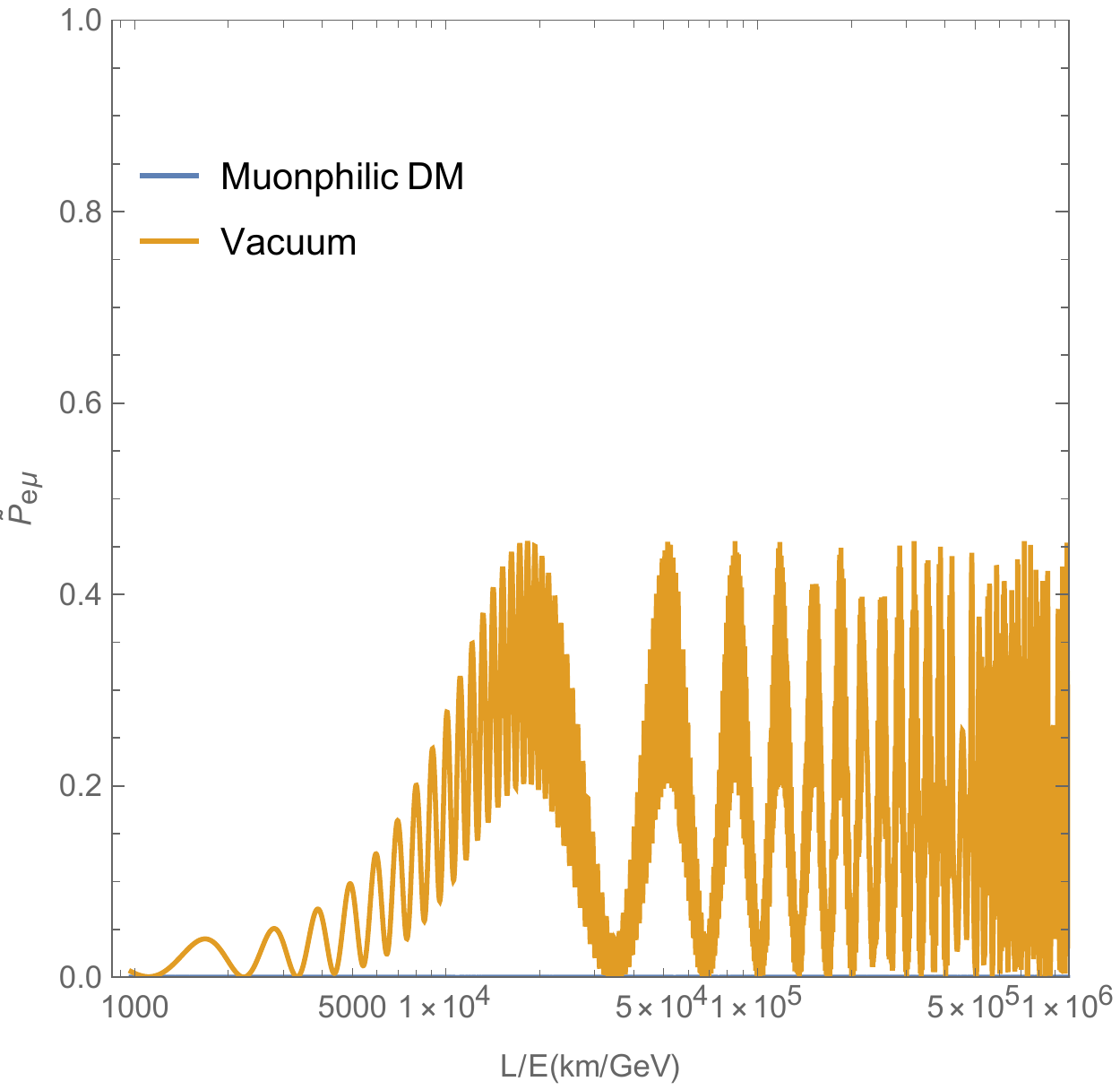}
\includegraphics[width=0.3\textwidth]{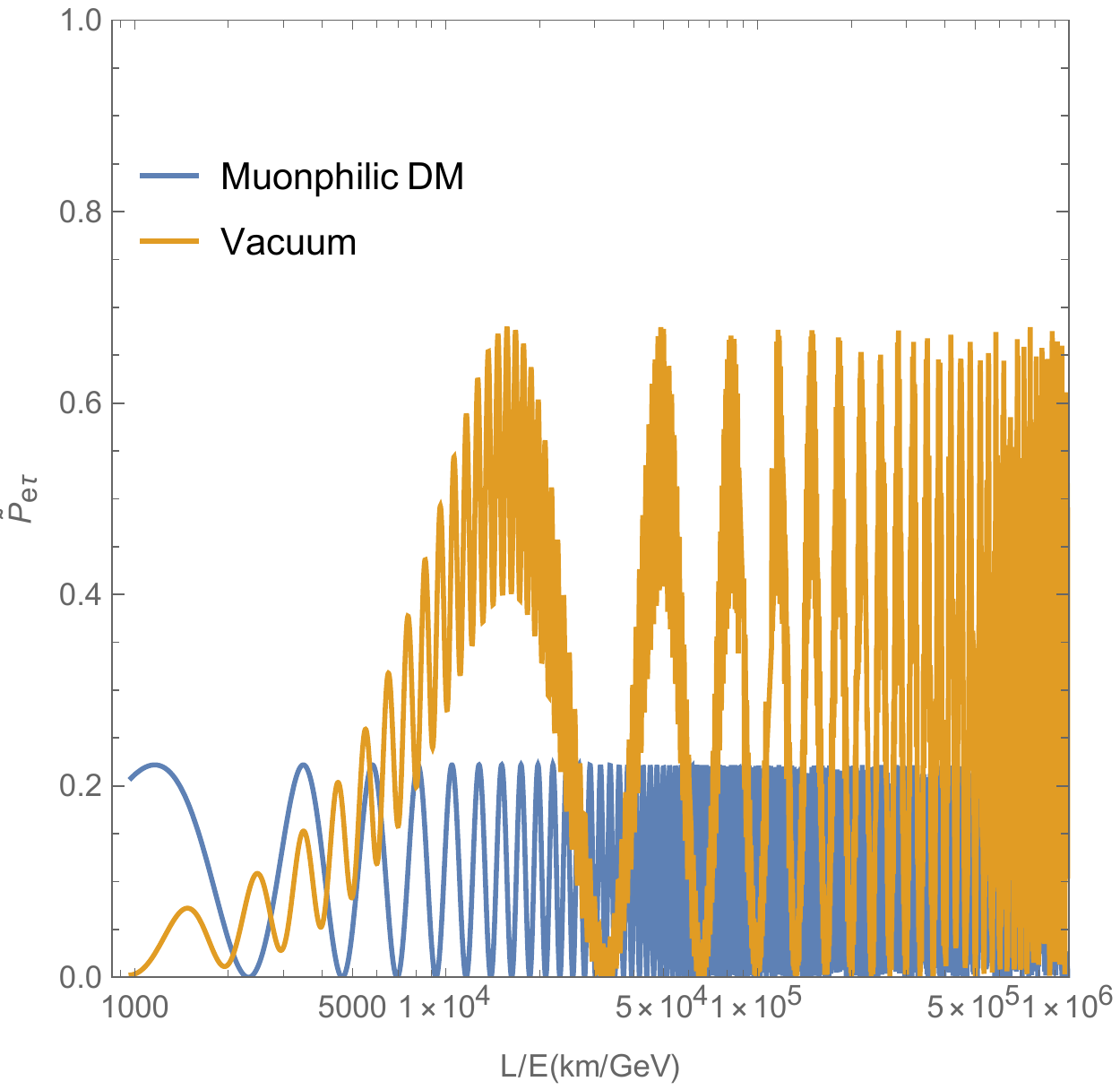} \\ 
\includegraphics[width=0.3\textwidth]{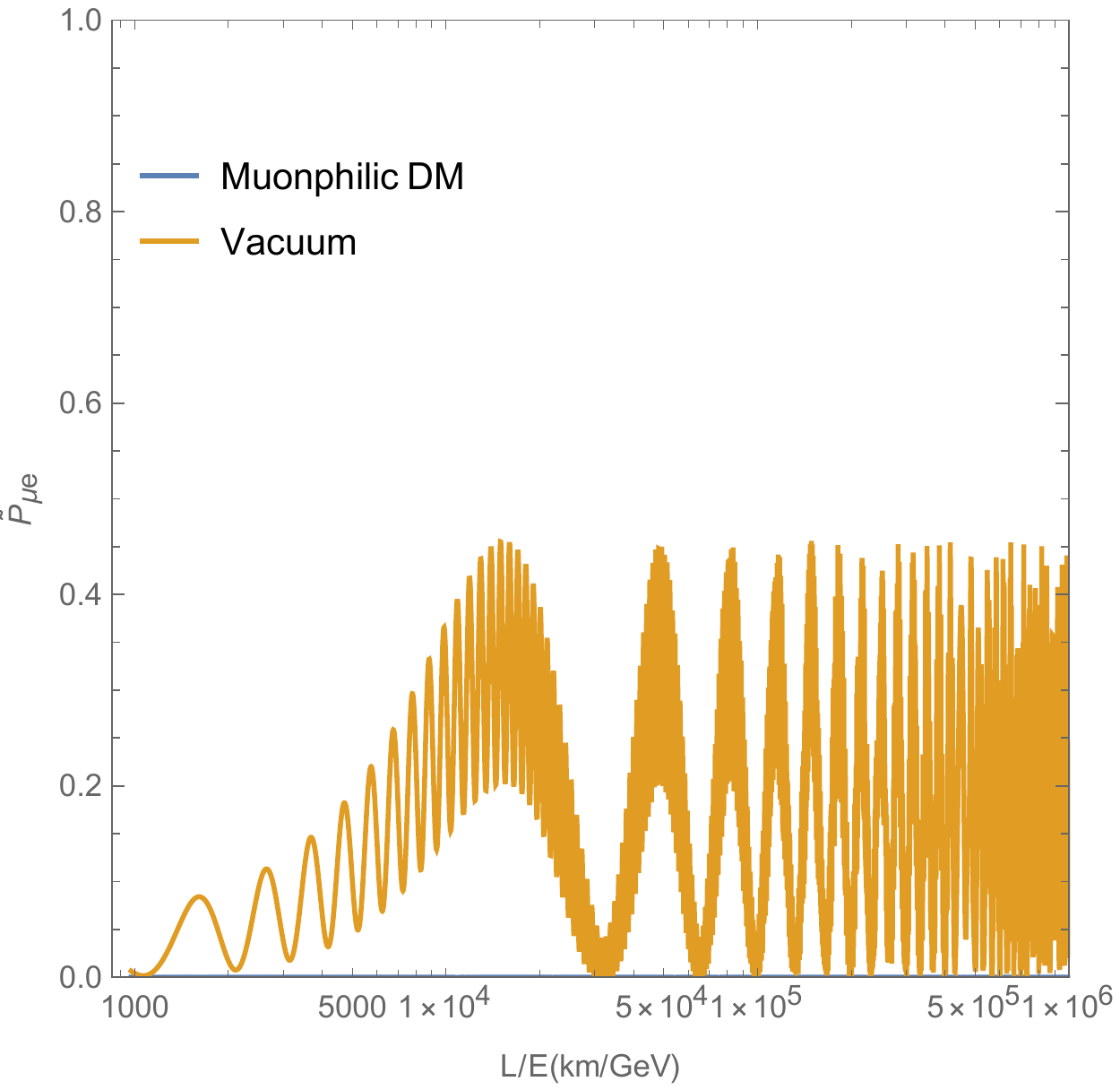} 
\includegraphics[width=0.3\textwidth]{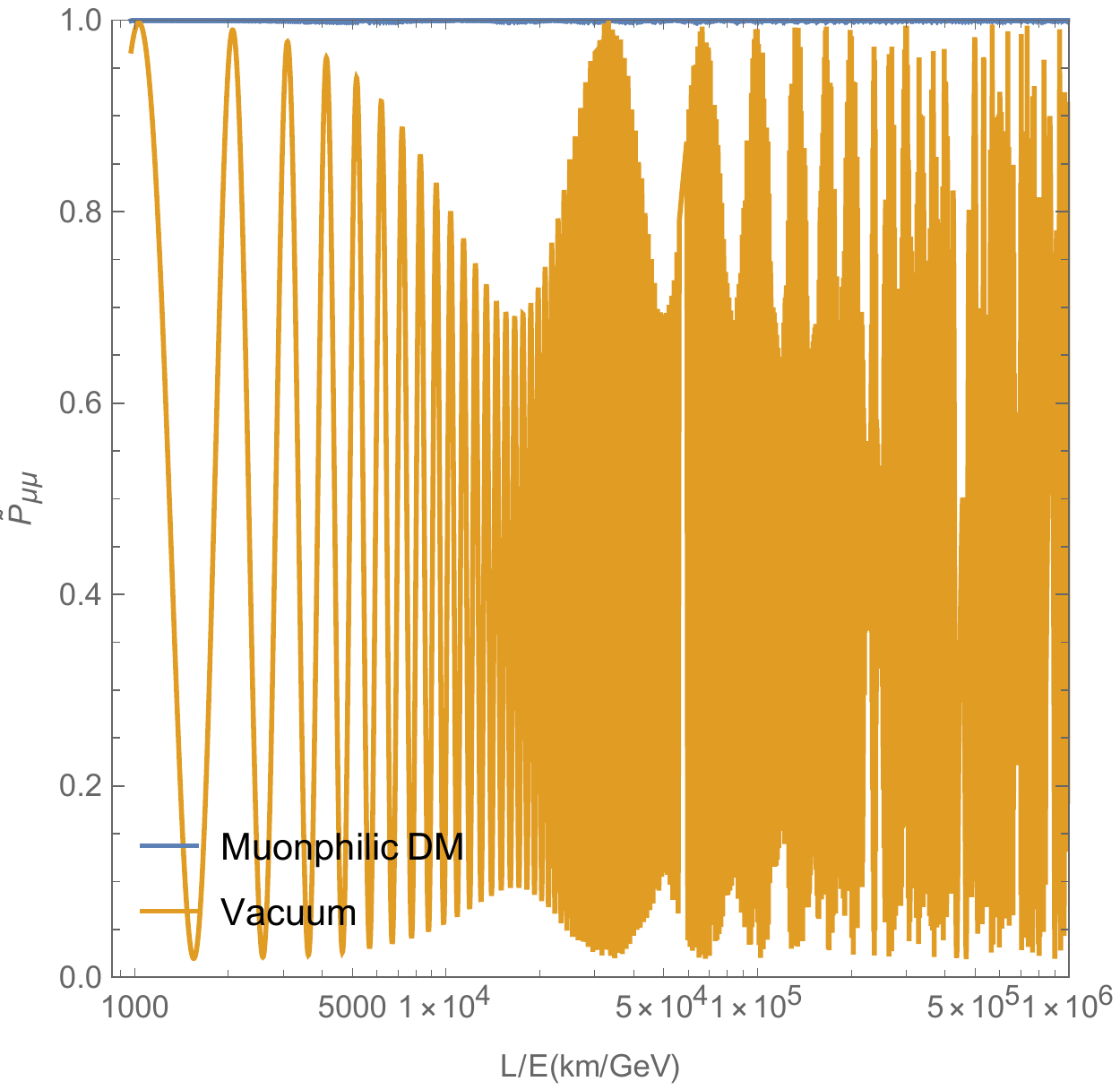}
\includegraphics[width=0.3\textwidth]{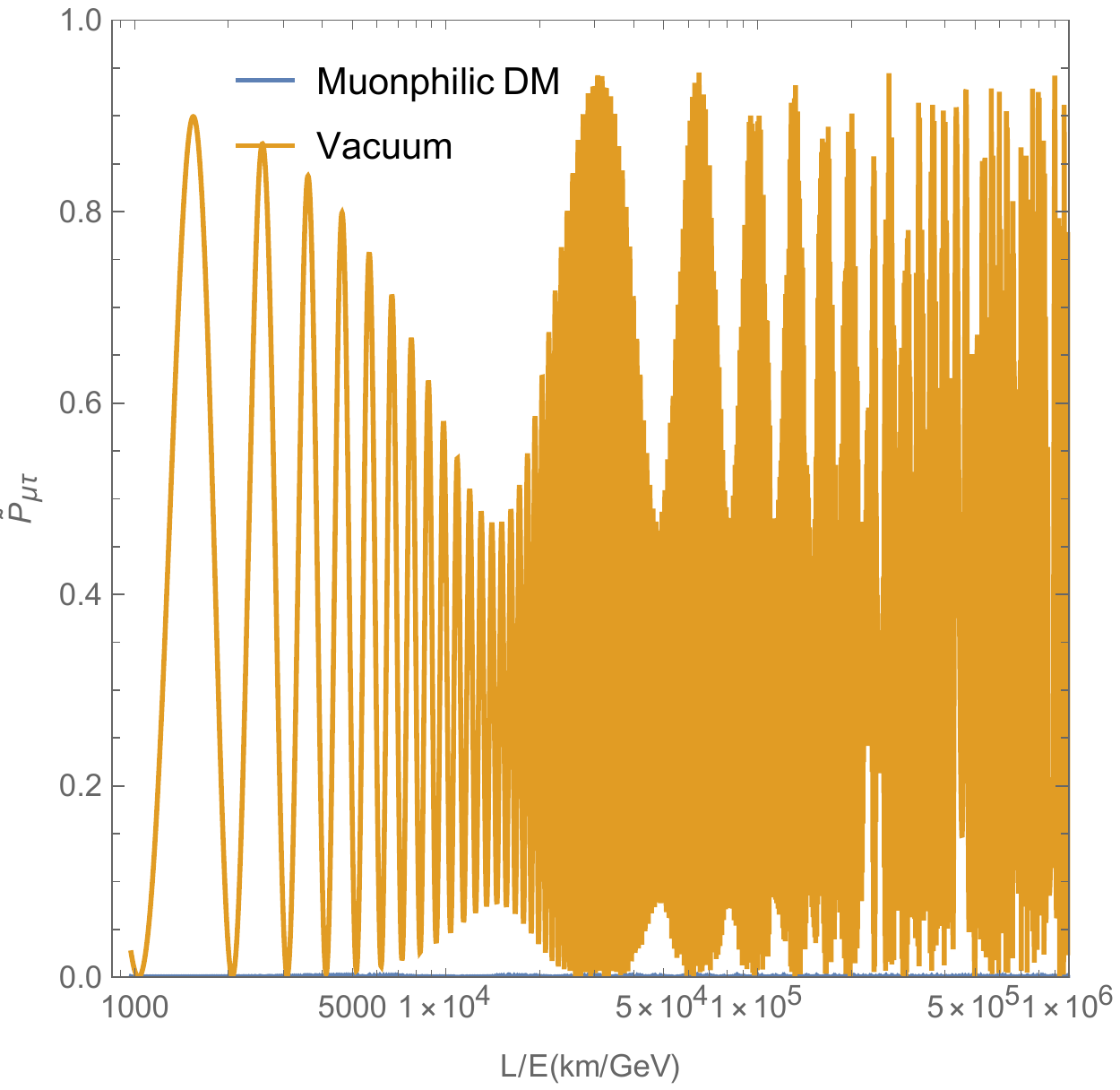}\\
\includegraphics[width=0.3\textwidth]{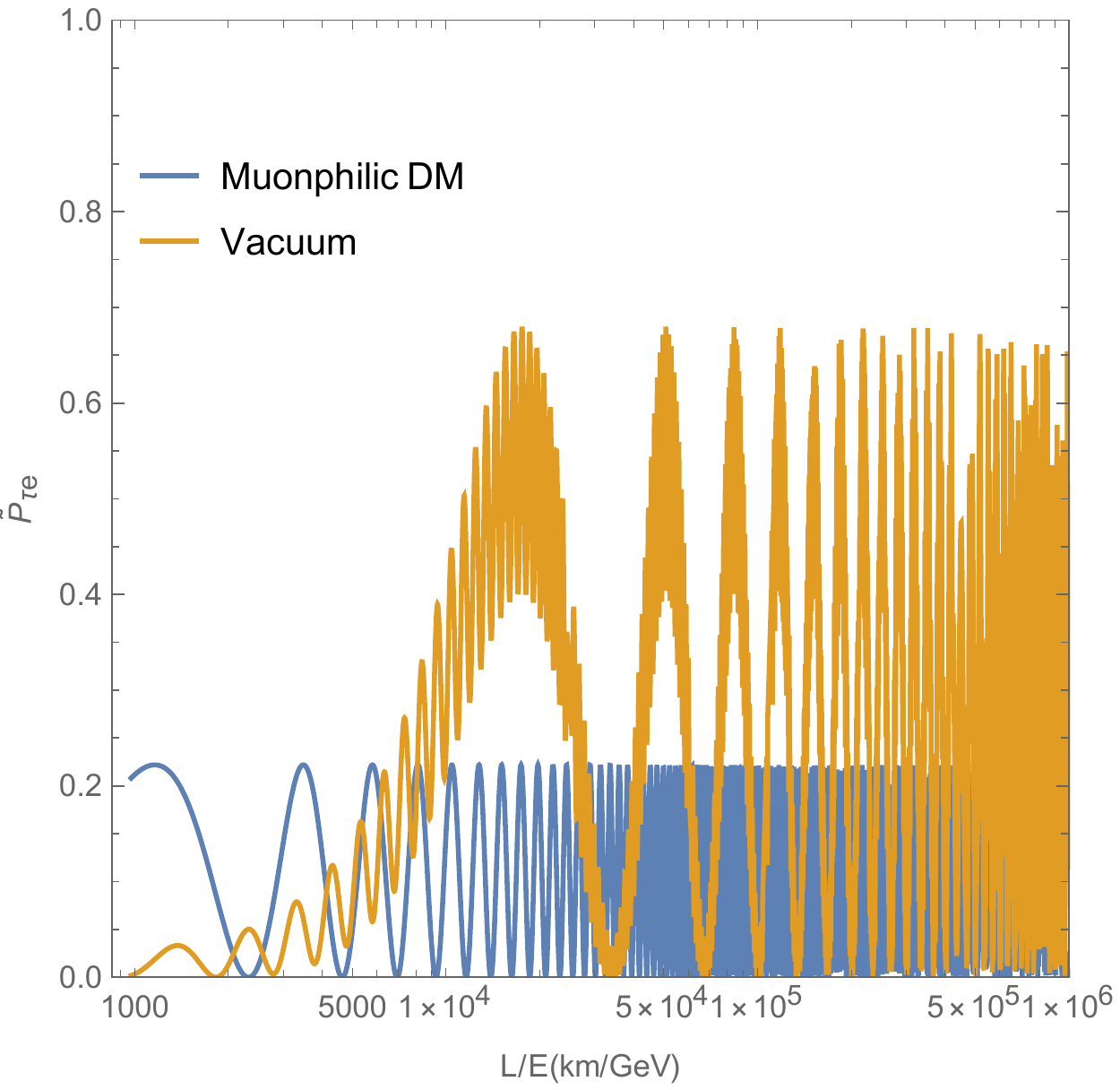} 
\includegraphics[width=0.3\textwidth]{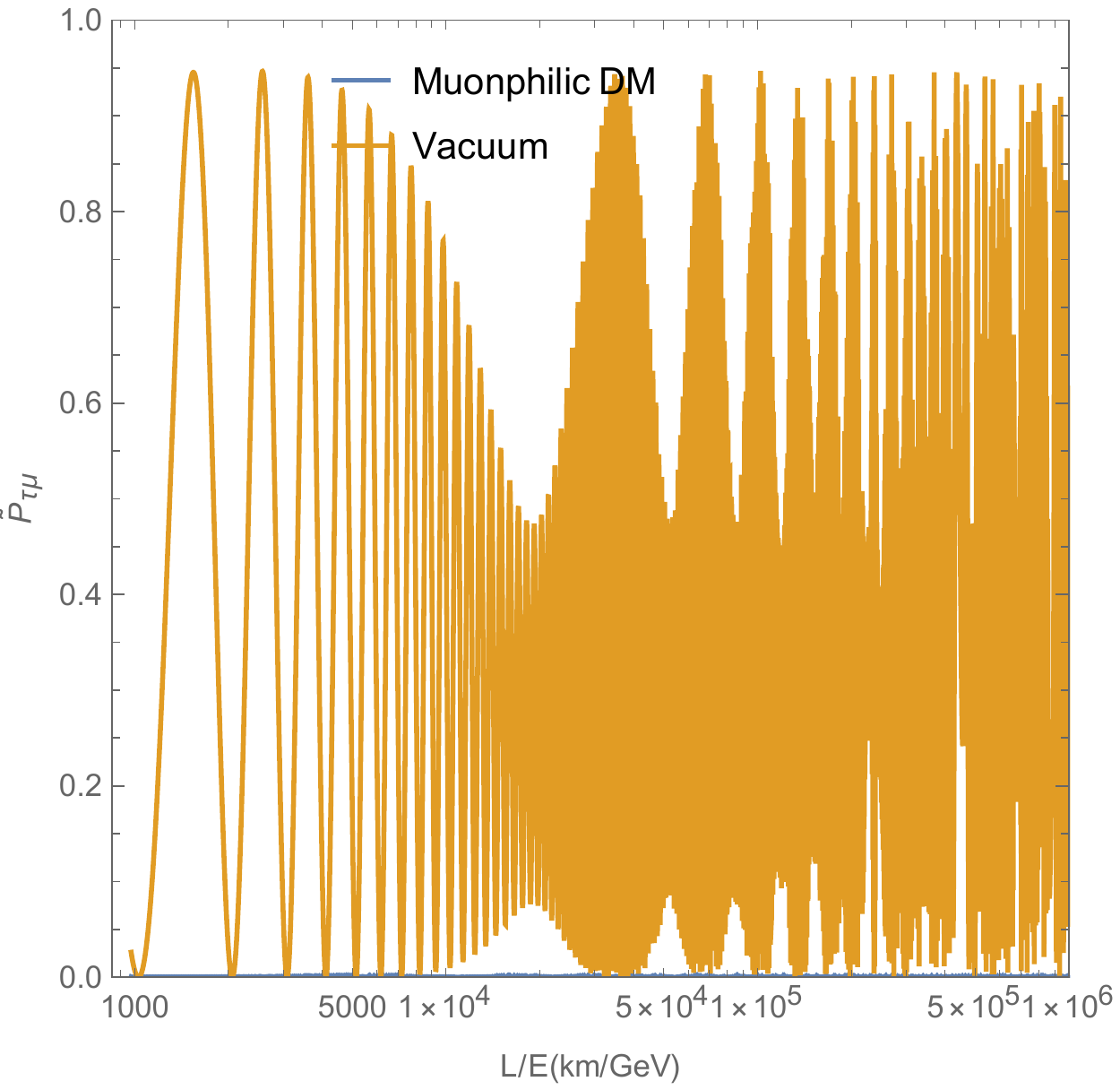}
\includegraphics[width=0.3\textwidth]{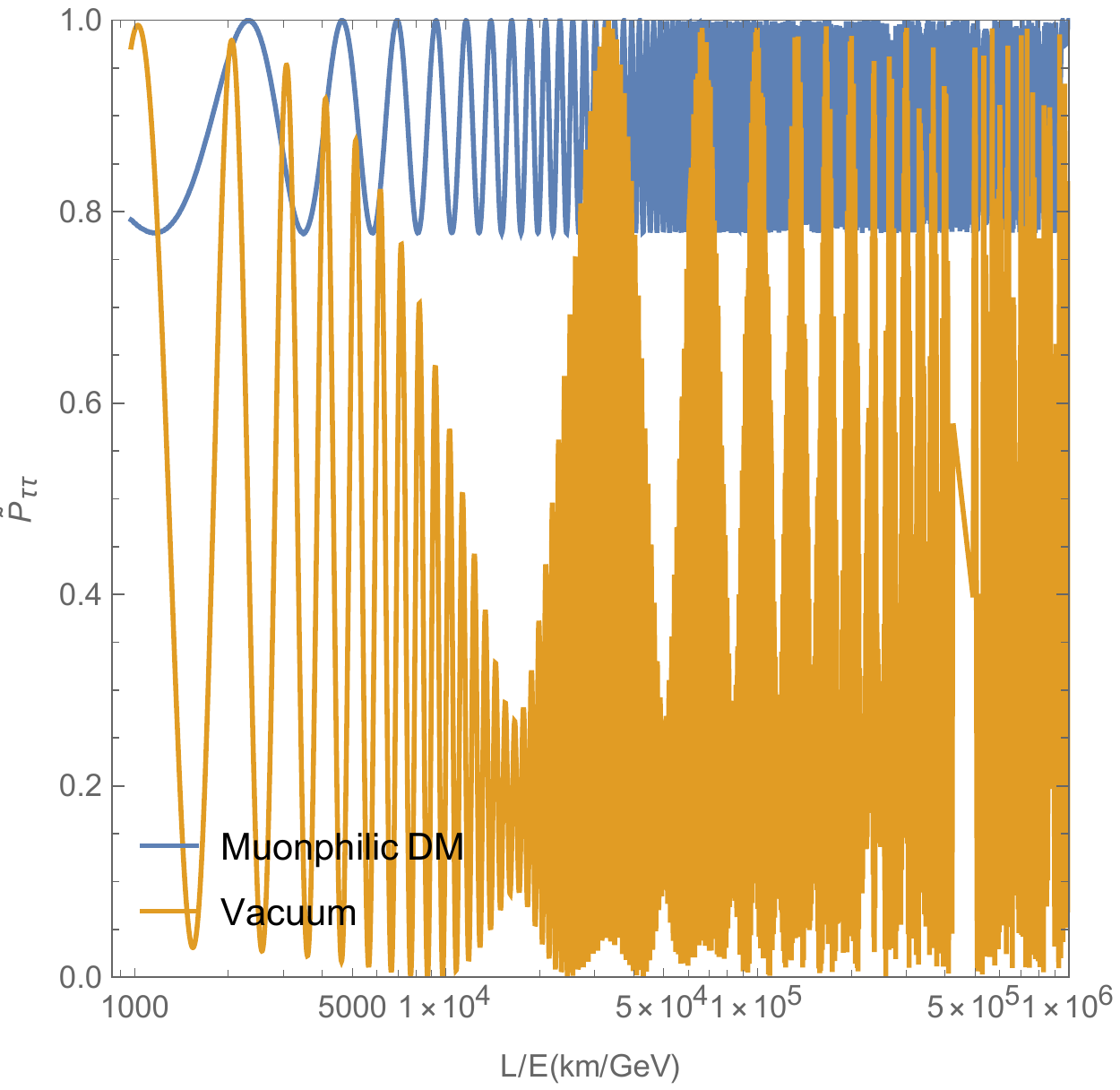}\\ 
\end{center}
\caption{Variation of neutrino survival and oscillation probabilities with $L/E$ for $A_{\chi}=0.1\, \rm eV^2$ (muonphilic DM).}
\label{le}
\end{figure*}

\section{Results and Discussions}
\label{results}

As discussed in Sec. \ref{formalism}, when neutrinos pass through a dense DM sub-halo, the oscillation probabilities would be affected by the DM potential. This effect however will be different for different two types of DM under consideration, as explained in Sec.~\ref{formalism}. This is due to the fact that the DM particle contributing to $b \to s \mu^+ \mu^-$, as introduced in Sec.~\ref{phi}, only interacts with muon neutrinos whereas the DM introduced in Sec.~\ref{z}, can couple to both $\nu_{\mu}$ and $\nu_{\tau}$.

Let us first consider the case of muonphilic DM \cite{Cerdeno:2019vpd}. The neutrino oscillation pattern as a function of $A_{\chi}$ is shown in Fig.~ \ref{nprob-1}. It is evident from Fig.~\ref{nprob-1} that the oscillation probabilities in DM appears like vacuum oscillations upto $A_{\chi}\sim 10^{-4}\, \rm  eV^{2}$. Above this value, oscillation probabilities in DM differ significantly from the vacuum scenario. Again above  $A_{\chi}\sim 10^{-2} \,\rm eV^{2}$, the survival probability of muon neutrinos approaches unity, i.e., the probability of oscillation from muon neutrino to other neutrino flavors and vice versa, becomes zero. Thus $\nu_{\mu}$ decouples from the other two neutrino flavors and only $\nu_{e}-\nu_{\tau}$ oscillation is possible above this region.

On the other hand, for DM charged under $L_{\mu}-L_{\tau}$, it was shown that  all the three flavors of neutrino decouple from each other  \footnote{For example, this is possible for $m_\chi = 10$ GeV, $m_{Z'} = 0.1$ GeV and $g_X = 0.1$ which corresponds to $A_\chi = {7.6}\,\rm eV^{2}$. These couplings and particle masses are consistent with the correct relic density \cite{Chao:2020qpe}.}\cite{Chao:2020qpe}.  A change in oscillation pattern can thus enable a discrimination between the two types of DM.  Thus one needs to analyze the oscillation probability patterns for high-$A_{\chi}$ values. According to eq.~\eqref{dmpot}, this requires very high number density of DM.   Such high number density is suggested  in some of the recent DM distribution models, for e.g., \cite{Arguelles:2016ihf,arguelles:2018novel}. 

A legitimate question arises whether the desired values of $A_{\chi}$ are attainable within the considered model. As one can readily see, the desired values of $A_{\chi}(=2EV_{\chi})$, can be achieved for UHE ($\sim$ 100 TeV - 1 PeV) neutrinos for the mass and coupling accommodating B-anomalies. For example, the best scenario to reconcile the B-anomalies is given by $m_{\chi} $ = 10 GeV and $m_{\phi}$ = 358 GeV for $\lambda_{\mu} = \sqrt{4\pi}$ which are also are consistent with the observed DM relic density \cite{Cerdeno:2019vpd}. For these values, the $A_{\chi}$ corresponding to 1 PeV neutrinos is $ 0.9\, \rm  eV^{2}$. This can be obtained for the DM density as suggested by the RAR model of DM ($\sim 10^{29}$ eV/cc). For 100 TeV neutrinos, $A_\chi$ is one order of magnitude lesser than the values corresponding to the 1 PeV neutrinos. 

 The number density of electrons and protons at the galactic center is $n_p \sim n_e \sim 10^4\,\rm{cc^{-1}}$ \cite{Fryer:2006wy,Herter:1984}. The protons and electrons exists in the DM halos in the form of gas molecules. This gives rise to the potential $V \sim 10^{-33} \rm{eV}$, which  for 1 PeV energetic neutrinos correspond to $A_\chi \sim 10^{-18}\, \rm eV^{2}$. The DM potential is many orders of magnitude larger than this SM potential. Thus we can safely neglect the SM potential in the context of this work.

The LHC dijet and dimuon searches are expected to put strong bounds on the allowed parameter space of this model. The colored as well as the leptonic scalar particles in this model can be pair produced at the ATLAS and CMS experiments. The colored scalars would eventually decay into quarks along with a DM particle which is identified from the missing transverse energy $\slashed{
{E_T}}$ whereas the decays of leptonic scalar can lead to $\mu\mu/\mu\nu+ \slashed{{E_T}}$. In \cite{Cerdeno:2019vpd} it was shown that the parameter space preferred by the flavor data is not excluded by the LHC constraints. The direct searches such as the Xenon1T results \cite{XENON:2018voc} are also expected to provide useful constraints. However, this model survives the current data. The next generation experiments with multi-ton targets, such as DarkSide \cite{DarkSide-20k:2017zyg}, are expected to probe this model in the mass range $m_{\chi}$ of (10-60) GeV.

We consider a cosmic  neutrino flux of  energy $E$ = 1 PeV passing through a muonphilic  DM  sub-halo of radius $10^{-3}$ pc  with  $A_{\chi}=0.1\,\rm eV^{2}$ and  situated near the center of the Milky Way \footnote{The radius of the halo is considered to be $10^{-3}$ pc, since a larger DM halo with such a high density is not allowed by the galaxy rotation curve \cite{Arguelles:2016ihf,arguelles:2018novel}. }. The variation of neutrino oscillation probabilities with $L/E$, for $A_{\chi} = 0.1\, \rm eV^{2}$, is shown in Fig.~\ref{le}. As one can see from Fig.~\ref{nprob-1}, this variation is valid for any higher values of $A_{\chi}$.   It is evident that this behaviour of oscillation pattern in the muonphilic DM departs from that of vacuum as well as DM charged under $L_{\mu}-L_{\tau}$.  This implies that, after traversing through the DM halo,  the flavor composition of neutrinos would be different from that of propagation through the free space. After emerging from the DM halo, the flavor composition of neutrinos can change further on their journey to the detector placed on the Earth. Therefore the effects of neutrinos and DM interactions resulting in the change of neutrino oscillation patterns can be detected on the Earth only if the flavor ratios are different from that of the vacuum oscillations.

This is due to the fact that, for a neutrino flux sourced at a very large distance from earth, the neutrino oscillation probabilities are averaged out. This generates a flavor ratio on earth, which deviates from its value at the source. Suppose the flux ratio of a particular flavor($\alpha$) of neutrino at the source is $\Phi_{\alpha}^{s}$ and the (averaged) oscillation probability for oscillation from flavor $\alpha$ to $\beta$ is given by $P_{\alpha\beta}$. The flux ratio of flavor $\beta$ on earth is then $\Phi_{\beta}^{\oplus} = \sum_{\alpha,\beta} P_{\alpha\beta} \Phi_{\alpha}^{s} $. This is what is expected when  flux of neutrino is propagating through vacuum. Now if the neutrino flux passes through a DM halo, the oscillation pattern will be different from that of vacuum as shown in Fig. \ref{le}.

In order to calculate the flavor ratios at earth, it is important to know the flavor components at the surface of the halo as after that it will be traveling through vacuum. We will consider this value as the  initial flux for averaging out through vacuum. Here we consider high energetic neutrino flux passing through a DM halo of radius $10^{-3}$ pc of DM potential $A_\chi = 0.1\,\rm eV^{2}$. As is evident from fig.\ref{le}, the neutrino in this situation only completes  a few cycles and hence the condition for averaging out is not satisfied within the DM sub-halo. Therefore, we consider, the flux corresponding to the probability at the surface of the sub-halo to be the initial flux for averaging out the probability.  Thus, when the flux comes out of the DM sub-halo, the flux ratio is different from the initial flux ratio as well as from the vacuum oscillations. Also, after emerging out of the DM sub-halo neutrinos traverse the rest path through vacuum. Thus, we can consider the flux ratio just outside the DM sub-halo as ``effective flux ratio at the source'' ($\Phi_{\alpha}^{s'}$). Now the flux ratio on earth is given as, $\Phi_{\beta}^{\oplus'} = \sum_{\alpha,\beta} P_{\alpha\beta} \Phi_{\alpha}^{s'} $.

 Assuming that the neutrinos are produced from pion decay so that the initial flux is $(\Phi_{e}^{s} : \Phi_{\mu}^{s} : \Phi_{\tau}^{s} ) = (0.33 : 0.66 : 0)$, the imprinted flux ratio of this neutrino flux on reaching earth turns out to be $(\Phi_{e}^{\oplus'} : \Phi_{\mu}^{\oplus'} : \Phi_{\tau}^{\oplus'} ) = (0.285 : 0.368 : 0.345)$. This flavor ratio at earth is obtained through two steps as mentioned earlier. At first, the neutrino flavor ratio at the surface of the DM sub-halo ($\Phi_{\alpha}^{h}$) is taken from the oscillatory curve (fig. \ref{le}). This flavor ratio acts as the ``effective flux ratio at the source''  ($\Phi_{\alpha}^{s'}$). Eventually this $\Phi_{\alpha}^{s'}$ gives the final flavor ratio at earth after averaging out. Here the flavor ratio at the surface of DM sub-halo ($\Phi_{\alpha}^{h}$) is given by the relation $\Phi_{\beta}^{h} = \sum_{\alpha,\beta} P_{\alpha\beta}^{h} \Phi_{\alpha}^{s}$, where $P_{\alpha\beta}^{h}$ is the oscillation probabilities at the surface of the sub-halo. In the present case the values of these probabilities are: ($P_{ee}^{h} = 0.854$, $P_{e \mu}^{h} = 2.2 \times 10^{-5}$, $P_{\mu \mu}^{h} = 0.999$, $P_{e \tau}^{h} = 0.145$, $P_{\mu \tau}^{h} = 3.5 \times 10^{-4}$), which are obtained from fig.\ref{le}. This  translates into the flavor ratio at the surface of the sub-halo as $(\Phi_{e}^{h} : \Phi_{\mu}^{h} : \Phi_{\tau}^{h} ) = ( 0.854 : 1.999  : 0.145 ) = (\Phi_{e}^{s'} : \Phi_{\mu}^{s'} : \Phi_{\tau}^{s'} )$ which  finally leads to the flavor ratio at earth as $(\Phi_{e}^{\oplus'} : \Phi_{\mu}^{\oplus'} : \Phi_{\tau}^{\oplus'} ) = (0.285 : 0.368 : 0.345)$ on averaging out. In the case of vacuum oscillations (without encountering DM sub-halo), the flux ratio is $(\Phi_{e}^{\oplus} : \Phi_{\mu}^{\oplus} : \Phi_{\tau}^{\oplus} ) = (0.308 : 0.351 : 0.339)$, which is calculated through \begin{equation}
    f_{\beta} = \sum_{\alpha}\left(\sum_{i}\abs{U_{\beta i}U_{\alpha i}^{*}}^{2} f_{\alpha}^{0}\right)\,,
    \label{reduced}
\end{equation}
where $f_{\alpha}^{0}$ is the initial flavor ratio of the flavor $\alpha$. This difference is, in principle, within the reach of the  future experimental precision \cite{Song:2020nfh}.
On the other hand, for the $L_{\mu}-L_{\tau}$ DM, since the three neutrino flavors are decoupled, the flux ratio on earth will resemble the case of vacuum oscillations. Therefore the interaction of neutrinos with dense sub-halo of DM has the potential to differentiate between the flavor models associated with DM particles.

Further, it would be interesting to see whether such a discrimination would be possible even for the sub-PeV flux. The fact that the number of observed PeV events are rare, a more decisive estimate of flavor compositions can be made by making use  of sub-PeV cosmic neutrino flux. 
If the energy of the neutrino flux is $E = 100$ \rm{TeV}, then $A_\chi$ reduces by one order of magnitude. We consider the same sub-halo size as for the PeV scenario. This changes ${L}/{E}$ to $10^5$, which modifies the oscillation probabilities. For this specification of energy and DM sub-halo size, the flavor ratio on earth turns out to be $(\Phi_{e}^{\oplus'} : \Phi_{\mu}^{\oplus'} : \Phi_{\tau}^{\oplus'} ) = (0.297 : 0.360 : 0.342)$. Indeed this flavor ratio is different from the expected flavor pattern corresponding to vacuum oscillations. Thus sub-PeV neutrinos can also play an important role in discriminating these types of DM.

The above conclusions are demonstrated assuming a profile with fixed dark matter density. We now consider the effects of density  variation. For this, we consider a phenomenological DM density profile, the Navarro-Frenk-White (NFW) model \cite{Navarro:1995iw}. The DM density in this model is given as
\begin{equation}
    \rho_{\chi} = \rho_{\oplus}\left(\frac{r_\oplus}{r}\right)\left(\frac{1 + \frac{r_\oplus}{r_s}}{1 + \frac{r}{r_s}}\right)^2,
    \label{NFW}
\end{equation}
where $r$ is the radial distance from the galactic center and  $r_\oplus = 8.5$ \rm{kpc} is the radial distance of the earth from the galactic center. Also, $r_s$ is a sub-halo specific parameter which for the Milky Way galaxy is 20 kpc. The local DM density is given by $\rho_\oplus = 0.4$ GeV/cc.  It can be observed from Eq.\ref{NFW} that the density of DM decreases in radially outward directions.
 For this DM distribution, the values of $A_\chi$ (E = 1 PeV) at the center of the sub-halo, for: (i) Muonphilic DM with mass $m_\chi = 10$ GeV is $A_\chi = 4.7 \times 10^{-14}\, \rm eV^{2}$ and for (ii) $L_\mu-L_\tau$ DM, with mass $m_\chi = 10$ \rm GeV is $A_\chi = 7.6 \times 10^{-9}\, \rm eV^{2}$. For 100 TeV, the value of $A_\chi$ decreases by 10 orders of magnitude.   

The DM candidates considered  here are massive ones possessing masses of the order of \rm{GeV}. The number density of DM is provided by, $n_\chi = \frac{\rho_\chi}{m_\chi}$. In the context of DM candidates considered here, this renders a very small DM number density, which in turn results in a tiny $A_\chi$. For these values of $A_\chi$ at any given point (even at the core), we find that the PMNS matrix elements resemble that in the vacuum. Since the flavor ratios depend on the PMNS matrix elements, they would remain unchanged for the NFW density profile.

We now consider another DM density profile, the isotropic DM density profile which suggests the DM density to be
\begin{equation}
    \rho_{\chi}  = \rho_{\oplus}\left(\frac{1 + (\frac{r_\oplus}{r_s})^2}{1 + (\frac{r}{r_s})^2}\right).
\end{equation}
In this profile, $r_s=5$ \rm{kpc}. Note that the cuspy core is absent in this profile. The number density of DM turns out to be very small in this profile as well. For isotropic DM distribution, the values of $A_\chi$ (E = 1 PeV) at the center of the sub-halo, for: (i) Muonphilic DM with mass $m_\chi = 10$ GeV is $A_\chi = 4.7 \times 10^{-23}\, \rm eV^{2}$ and for (ii) $L_\mu-L_\tau$ DM, with mass $m_\chi = 10$ GeV  is $A_\chi =7.6 \times 10^{-18}\, \rm eV^{2}  $. Thus it can be seen that for both NFW and isotropic  DM distributions, the DM potential is much lower than the $A_\chi$ for which it can alter the PMNS matrix elements (fig. \ref{nprob-1}). This is valid for both types of DM models. Thus it is expected that the oscillation pattern as well as the flavor ratios would be the same as for the vacuum oscillations.

\begin{figure}
    \centering
    \includegraphics[width=0.5\textwidth]{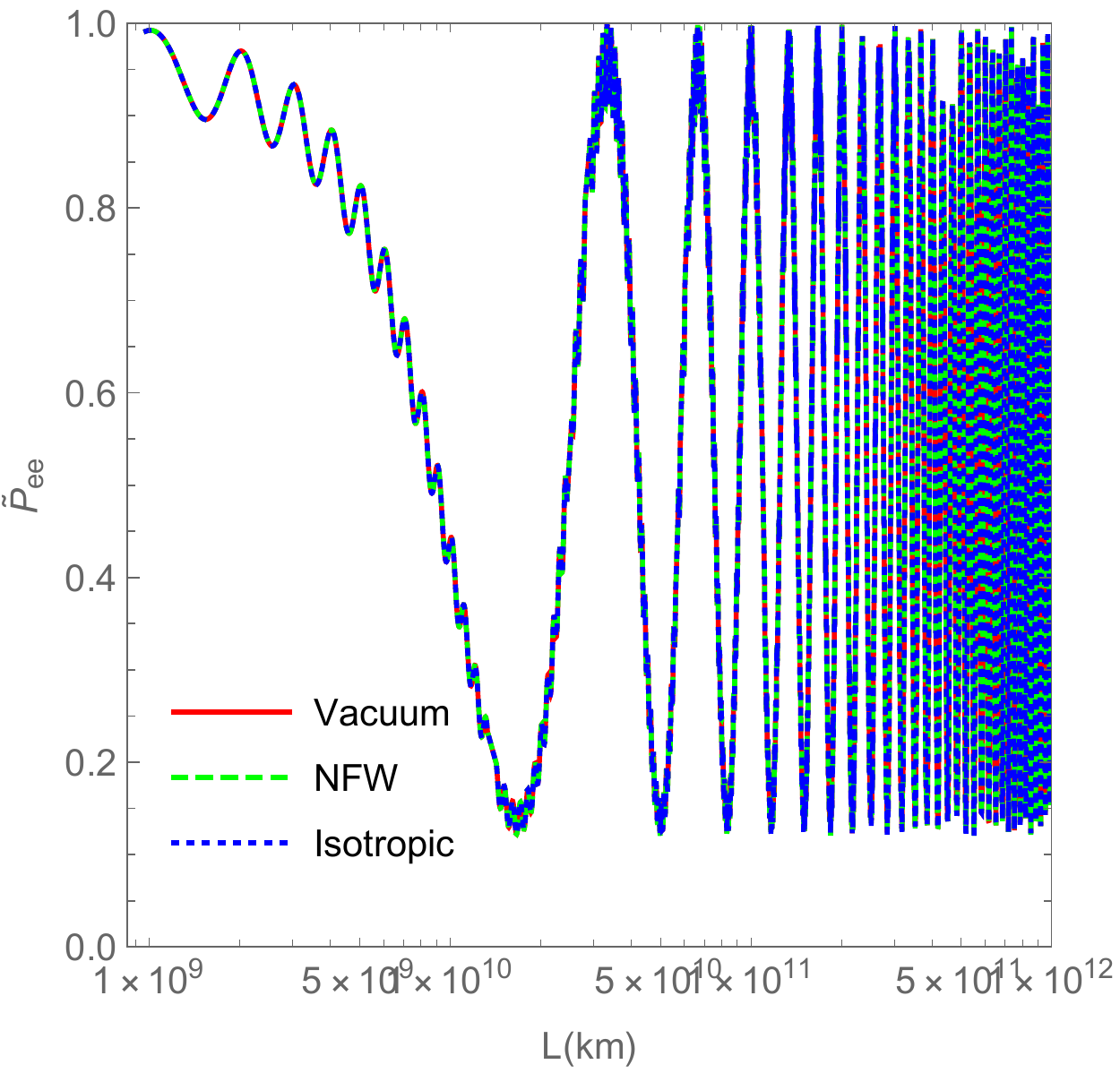}
    \caption{Comparison of $\Tilde{P}_{ee}$ in the NFW and isotropic DM profiles with vacuum (for E = 1 PeV)}
    \label{dmvac}
\end{figure}

We now show this by comparing the oscillation probabilities for both the NFW and isotropic DM density profiles with that of the vacuum oscillations in fig.\ref{dmvac}.  Here $\tilde{P}_{ee}$ for both the DM profiles is calculated by making use of the variation of $A_{\chi}$ with $L$ for E = 1 PeV.  From the figure, it is evident that these patterns coincide with that of the vacuum oscillations.
The evolution of flavor states in a medium with varying density is given as \cite{deSalas:2016svi}
\begin{equation}
    f_{\beta}(l_{n+1},\phi) = \sum_{\alpha}\left(\sum_{i}\abs{U_{\beta i}(l_{n},\phi)U_{\alpha i}^{*}(l_{n},\phi)}^{2} f_{\alpha}(l_{n},\phi)\right),
    \label{vary}
\end{equation}
where $l$ is the distance to the source and $\phi$ is the angle between the distance to the source and the straight line between earth and galactic center. 
As seen from fig. \ref{nprob-1} as well as fig. \ref{dmvac}, the PMNS matrix elements are unaltered for the two types of profiles. Thus \eqref{vary} for the final flavor states will reduce to \eqref{reduced}
which are the flavor states for propagation through vacuum over a large distance. We have also calculated the flavor ratios  using \eqref{vary}, i.e., with the varying densities for both the profiles and found that they coincide with that given by \eqref{reduced}.
For a significantly large $A_\chi$ required to modify the flavor patterns of neutrinos, a much larger DM number density is expected. This can be possible only if the DM follows the RAR-type of DM density distribution.

We now see whether the flavor composition changes for RAR-type of DM density profile. The RAR DM model takes into account keV-sterile neutrinos as DM particles \cite{arguelles:2018novel}. Unlike the other DM models, this model leads to a very dense matter concentrations near the core which may relinquish  flavor ratios different from vacuum oscillations. Although the DM candidates considered in this work are different in nature from the RAR model, the density of DM is adapted since it is allowed by the rotation curve of the galaxy. Owing to the radial dependence of DM density, the neutrino faces different strengths of potential at different points along the traversed path. This leads to distinct PMNS matrix elements at different points along the path.  Since the flavor ratios  depend on the PMNS matrix elements at different points, the flavor ratios changes accordingly. As explained before, the final flavor ratio at earth can be calculated using \eqref{vary}. We find that the flavor ratios of neutrinos on earth for the initial flavor ratio of (1 : 2 : 0) with this profile turns out to be (0.332 : 0.334 : 0.333) for 1 PeV neutrinos which is indeed different from the vacuum oscillations.

Lastly, we also take into account the consideration of the fact that the neutrino events can originate from more than one sources located at different places within the sub-halo. In the above analysis, the source was assumed to be located at the center of the sub-halo of radius $10^{-3}$ pc. We have also checked for the scenario where the neutrino flux is generated from a source which is located away from the sub-halo center  so that it traverses a lesser distance, say $10^{-4}$ pc, in the DM sub-halo.  As one can infer from Table.\ref{diff_dist}, the flavor ratios are different from vacuum both for PeV and sub-PeV events.

\begin{table}
\begin{centering}
\begin{tabular}{|c|c|c|}
     \hline
     Energy (E) &  Distance (in pc) & Flavor ratio \\
     \hline
     \multirow{3}{4em}{1 PeV} & $10^{-3}$ & (0.285 : 0.368 : 0.345) \\                                & $10^{-4}$ & (0.279 : 0.372 : 0.347) \\
     \hline            
     \multirow{3}{4em}{100 TeV} & $10^{-3}$ & (0.297 : 0.360 : 0.342) \\
                                & $10^{-4}$ & (0.293 : 0.362 : 0.343) \\
     \hline                       
\end{tabular}
\caption{Different distances traversed by the neutrinos in the DM sub-halo and the corresponding flavor ratios at the earth.}
\label{diff_dist}
\end{centering}
\end{table}

\section{Conclusion}
\label{conc}
The currently running experiments at the LHC have provided several enticing hints of physics beyond SM. This is corroborated by the combined measurements of muon $(g-2)$ at Fermilab and BNL which also disagrees with the SM prediction.  These anomalies can be attuned in a plenitude of models. A number of these models cognate to the dark sector. In this work, we consider two such models and study the impact of neutrino-dark matter interaction on the oscillation patterns of high energy cosmic neutrinos passing through dense dark matter sub-halo emplaced near the center of the Milky Way. 

For a model which incorporates a Majorana fermion which is a dark matter candidate along with  two new scalar fields and contributes to $b \to s$ decays at the one loop level, we find that the neutrino oscillation pattern after traversing through the DM sub-halo differs from that of vacuum oscillations. This would result in a different flavor ratio on the Earth as compared to free space oscillations. On the other hand, it was previously shown that for  $Z'$ model driven by $L_{\mu}-L_{\tau}$ symmetry which can accommodate the measurement of muon $(g-2)$ and also includes a vector-like fermion as a dark matter particle, the three flavors of neutrinos decouple from each other. This relinquishes a flavor ratio similar to that of vacuum oscillations. After reanalyzing with the updated measurement of muon $(g-2)$, the conclusions for this model remain unchanged. Therefore interaction of ultra high energy cosmic muon neutrinos with a dense sub-halo of dark matter has the potential to be  a good discriminant of new physics models which accommodate current anomalies in $b \to s \mu^+ \mu^-$ sector and/or muon $(g-2)$ and having a liaison with the dark sector.

\bigskip
\noindent
{\bf Acknowledgements}: The work of A.K.A. is supported by SERB-India Grant CRG/2020/004576. We would like to acknowledge Shireen Gangal for useful suggestions.
We thank anonymous referee for his/her constructive comments and suggestions for the improvement of our manuscript.


\begin{thebibliography}{99}

\bibitem{LHCb:2021trn}
R.~Aaij \textit{et al.} [LHCb],
Nature Phys. \textbf{18}, no.3, 277-282 (2022)
[arXiv:2103.11769 [hep-ex]].

\bibitem{rkstar}
R.~Aaij \textit{et al.} [LHCb],
JHEP \textbf{08} (2017), 055
[arXiv:1705.05802 [hep-ex]].

\bibitem{Hiller:2003js} 
  G.~Hiller and F.~Kruger,
  Phys.\ Rev.\ D {\bf 69}, 074020 (2004)
  [hep-ph/0310219].
  
\bibitem{Bordone:2016gaq} 
M.~Bordone, G.~Isidori and A.~Pattori,
  Eur.\ Phys.\ J.\ C {\bf 76}, no. 8, 440 (2016)
  [arXiv:1605.07633 [hep-ph]].

\bibitem{Isidori:2020acz}
G.~Isidori, S.~Nabeebaccus and R.~Zwicky,
JHEP \textbf{12} (2020), 104
[arXiv:2009.00929 [hep-ph]].

  \bibitem{Straub:2018kue} 
  D.~M.~Straub,
  arXiv:1810.08132 [hep-ph].
  
   

    \bibitem{sm-angular} 
      S.~Descotes-Genon, T.~Hurth, J.~Matias and J.~Virto,
  JHEP {\bf 1305}, 137 (2013)
  [arXiv:1303.5794 [hep-ph]].
  
  \bibitem{Kstarlhcb1}
R.~Aaij {\it et al.} [LHCb Collaboration],
  Phys.\ Rev.\ Lett.\  {\bf 111}, 191801 (2013)
  [arXiv:1308.1707 [hep-ex]].
  
  
    \bibitem{Kstarlhcb2}
R.~Aaij {\it et al.} [LHCb Collaboration],
  JHEP {\bf 1602}, 104 (2016)
  [arXiv:1512.04442 [hep-ex]].
  
    \bibitem{bsphilhc2}
R.~Aaij {\it et al.} [LHCb Collaboration],
  JHEP {\bf 1509}, 179 (2015)
  [arXiv:1506.08777 [hep-ex]].
  
  
  
       
  \bibitem{Descotes-Genon:2013wba}
S.~Descotes-Genon, J.~Matias and J.~Virto,
Phys. Rev. D \textbf{88}, 074002 (2013)
[arXiv:1307.5683 [hep-ph]].

\bibitem{Altmannshofer:2013foa}
W.~Altmannshofer and D.~M.~Straub,
Eur. Phys. J. C \textbf{73}, 2646 (2013)
[arXiv:1308.1501 [hep-ph]].



\bibitem{Alok:2017jgr}
A.~K.~Alok, B.~Bhattacharya, D.~Kumar, J.~Kumar, D.~London and S.~U.~Sankar,
Phys. Rev. D \textbf{96} (2017) no.1, 015034
[arXiv:1703.09247 [hep-ph]].

 
 \bibitem{Alok:2019ufo}
A.~K.~Alok, A.~Dighe, S.~Gangal and D.~Kumar,
JHEP \textbf{06} (2019), 089
[arXiv:1903.09617 [hep-ph]].


  \bibitem{Altmannshofer:2021qrr}
W.~Altmannshofer and P.~Stangl,
Eur. Phys. J. C \textbf{81}, no.10, 952 (2021)
[arXiv:2103.13370 [hep-ph]].
  


\bibitem{Carvunis:2021jga}
A.~Carvunis, F.~Dettori, S.~Gangal, D.~Guadagnoli and C.~Normand,
[arXiv:2102.13390 [hep-ph]]

\bibitem{Alguero:2021anc}
M.~Alguer\'o, B.~Capdevila, S.~Descotes-Genon, J.~Matias and M.~Novoa-Brunet,
[arXiv:2104.08921 [hep-ph]]

\bibitem{Geng:2021nhg}
L.~S.~Geng, B.~Grinstein, S.~J\"ager, S.~Y.~Li, J.~Martin Camalich and R.~X.~Shi,
[arXiv:2103.12738 [hep-ph]]

\bibitem{Hurth:2021nsi}
T.~Hurth, F.~Mahmoudi, D.~M.~Santos and S.~Neshatpour,
[arXiv:2104.10058 [hep-ph]].


\bibitem{Angelescu:2021lln}
A.~Angelescu, D.~Be\v{c}irevi\'c, D.~A.~Faroughy, F.~Jaffredo and O.~Sumensari,
Phys. Rev. D \textbf{104}, no.5, 055017 (2021)
[arXiv:2103.12504 [hep-ph]].

\bibitem{Alok:2022pjb}
A.~K.~Alok, N.~R.~Singh Chundawat, S.~Gangal and D.~Kumar,
Eur. Phys. J. C \textbf{82}, no.10, 967 (2022)
[arXiv:2203.13217 [hep-ph]].

\bibitem{SinghChundawat:2022zdf}
N.~R.~Singh Chundawat,
[arXiv:2207.10613 [hep-ph]].

   \bibitem{Abi:2021gix}
B.~Abi \textit{et al.} [Muon g-2],
Phys. Rev. Lett. \textbf{126} (2021) no.14, 141801
[arXiv:2104.03281 [hep-ex]].

\bibitem{Bennett:2006fi}
G.~W.~Bennett \textit{et al.} [Muon g-2],
Phys. Rev. D \textbf{73} (2006), 072003
[arXiv:hep-ex/0602035 [hep-ex]].


\bibitem{Vicente:2018frk}
A.~Vicente,
Springer Proc. Phys. \textbf{234}, 393-400 (2019)
[arXiv:1812.03028 [hep-ph]].

\bibitem{London:2021lfn}
D.~London and J.~Matias,
[arXiv:2110.13270 [hep-ph]].

\bibitem{Gandhi:1998ri}
R.~Gandhi, C.~Quigg, M.~H.~Reno and I.~Sarcevic,
Phys. Rev. D \textbf{58}, 093009 (1998)
[arXiv:hep-ph/9807264 [hep-ph]].

\bibitem{Bhattacharya:2010xj}
A.~Bhattacharya, S.~Choubey, R.~Gandhi and A.~Watanabe,
JCAP \textbf{09}, 009 (2010)
[arXiv:1006.3082 [hep-ph]].


\bibitem{Hooper:2004xr}
D.~Hooper, D.~Morgan and E.~Winstanley,
Phys. Lett. B \textbf{609}, 206-211 (2005)
[arXiv:hep-ph/0410094 [hep-ph]].

\bibitem{Ackermann:2022rqc}
M.~Ackermann, S.~K.~Agarwalla, J.~Alvarez-Mu\~niz, C.~A.~Arg\"uelles, M.~Bustamante, B.~A.~Clark, A.~Cummings, V.~Decoene, P.~B.~Denton and D.~Dornic, \textit{et al.}
[arXiv:2203.08096 [hep-ph]].


\bibitem{Farzan:2018pnk}
Y.~Farzan and S.~Palomares-Ruiz,
Phys. Rev. D \textbf{99}, no.5, 051702 (2019)
[arXiv:1810.00892 [hep-ph]].

\bibitem{Farzan:2021gbx}
Y.~Farzan,
JHEP \textbf{07}, 174 (2021)
[arXiv:2105.03272 [hep-ph]].

\bibitem{Farzan:2021slf}
Y.~Farzan,
PoS \textbf{EPS-HEP2021}, 261 (2022)
[arXiv:2110.07222 [hep-ph]].

\bibitem{Pandey:2018wvh}
S.~Pandey, S.~Karmakar and S.~Rakshit,
JHEP \textbf{01} (2019), 095
[erratum: JHEP \textbf{11} (2021), 215]
[arXiv:1810.04203 [hep-ph]].


\bibitem{Karmakar:2020yzn}
S.~Karmakar, S.~Pandey and S.~Rakshit,
JHEP \textbf{10} (2021), 004
[arXiv:2010.07336 [hep-ph]].

\bibitem{Brdar:2017kbt}
V.~Brdar, J.~Kopp, J.~Liu, P.~Prass and X.~P.~Wang,
Phys. Rev. D \textbf{97} (2018) no.4, 043001
[arXiv:1705.09455 [hep-ph]].

\bibitem{IceCube:2013cdw}
M.~G.~Aartsen \textit{et al.} [IceCube],
Phys. Rev. Lett. \textbf{111} (2013), 021103
[arXiv:1304.5356 [astro-ph.HE]].



\bibitem{Anchordoqui:2013dnh}
L.~A.~Anchordoqui, V.~Barger, I.~Cholis, H.~Goldberg, D.~Hooper, A.~Kusenko, J.~G.~Learned, D.~Marfatia, S.~Pakvasa and T.~C.~Paul, \textit{et al.}
JHEAp \textbf{1-2}, 1-30 (2014)
[arXiv:1312.6587 [astro-ph.HE]].

\bibitem{ANTARES:2011hfw}
M.~Ageron \textit{et al.} [ANTARES],
Nucl. Instrum. Meth. A \textbf{656} (2011), 11-38
[arXiv:1104.1607 [astro-ph.IM]].

\bibitem{KM3Net:2016zxf}
S.~Adrian-Martinez \textit{et al.} [KM3Net],
J. Phys. G \textbf{43} (2016) no.8, 084001
[arXiv:1601.07459 [astro-ph.IM]].

\bibitem{Avrorin:2022lyk}
A.~V.~Avrorin, A.~D.~Avrorin, V.~M.~Ayinutdinov, V.~A.~Allakhverdyan, P.~Banash, Z.~Bardachova, I.~A.~Belolaptikov, I.~V.~Borina, V.~B.~Brudanin and N.~M.~Budnev, \textit{et al.}
J. Exp. Theor. Phys. \textbf{134} (2022) no.4, 399-416

\bibitem{P-ONE:2020ljt}
M.~Agostini \textit{et al.} [P-ONE],
Nature Astron. \textbf{4}, no.10, 913-915 (2020)
[arXiv:2005.09493 [astro-ph.HE]].

\bibitem{Resconi:2021ezb}
E.~Resconi [P-ONE],
PoS \textbf{ICRC2021}, 024 (2022)
[arXiv:2111.13133 [astro-ph.IM]].

\bibitem{IceCube-Gen2:2020qha}
M.~G.~Aartsen \textit{et al.} [IceCube-Gen2],
J. Phys. G \textbf{48}, no.6, 060501 (2021)
[arXiv:2008.04323 [astro-ph.HE]].

\bibitem{IceCube:2014gqr}
M.~G.~Aartsen \textit{et al.} [IceCube],
[arXiv:1412.5106 [astro-ph.HE]].

\bibitem{Brown:2021lef}
A.~M.~Brown, M.~Bagheri, M.~Doro, E.~Gazda, D.~Kieda, C.~Lin, Y.~Onel, N.~Otte, I.~Taboada and A.~Wang,
[arXiv:2109.03125 [astro-ph.IM]].

\bibitem{RadarEchoTelescope:2021rca}
S.~Prohira \textit{et al.} [Radar Echo Telescope],
Phys. Rev. D \textbf{104}, no.10, 102006 (2021)
[arXiv:2104.00459 [astro-ph.IM]].

\bibitem{Romero-Wolf:2020pzh}
A.~Romero-Wolf, J.~Alvarez-Mu\~niz, W.~R.~Carvalho, A.~Cummings, H.~Schoorlemmer, S.~Wissel, E.~Zas, C.~Arg\"uelles, H.~Barreda and J.~Bazo, \textit{et al.}
[arXiv:2002.06475 [astro-ph.IM]].


\bibitem{penacchioni:2020testing}
A.~V.~Penacchioni, O.~Civitarese and C.~R.~Arg\"uelles,
Eur. Phys. J. C \textbf{80} (2020) no.3, 183


\bibitem{Navarro:1995iw}
J.~F.~Navarro, C.~S.~Frenk and S.~D.~M.~White,
Astrophys. J. \textbf{462}, 563-575 (1996)
[arXiv:astro-ph/9508025 [astro-ph]].

\bibitem{Navarro:1996gj}
J.~F.~Navarro, C.~S.~Frenk and S.~D.~M.~White,
Astrophys. J. \textbf{490}, 493-508 (1997)
[arXiv:astro-ph/9611107 [astro-ph]].

\bibitem{Sofue:2011kw}
Y.~Sofue,
Publ. Astron. Soc. Jap. \textbf{64}, 75 (2012)
[arXiv:1110.4431 [astro-ph.GA]].

\bibitem{Einasto:1965czb}
J.~Einasto,
Trudy Astrofizicheskogo Instituta Alma-Ata \textbf{5}, 87-100 (1965)

\bibitem{iso}
J.~Binney, S.~Tremaine, {\it Galactic Dynamics}, 2nd edn., Princeton University Press, Princeton, 2008.

\bibitem{Arguelles:2016ihf}
C.~R.~Arg\"uelles, A.~Krut, J.~A.~Rueda and R.~Ruffini,
Phys. Dark Univ. \textbf{21}, 82-89 (2018)
[arXiv:1606.07040 [astro-ph.GA]].

\bibitem{arguelles:2018novel}
C.~R.~Arg\"uelles, A.~Krut, J.~A.~Rueda and R.~Ruffini,
Phys. Dark Univ. \textbf{24}, 100278 (2019)
[arXiv:1810.00405 [astro-ph.GA]].

\bibitem{Cerdeno:2019vpd}
D.~G.~Cerde\~no, A.~Cheek, P.~Mart\'\i{}n-Ramiro and J.~M.~Moreno,
Eur. Phys. J. C \textbf{79}, no.6, 517 (2019)
[arXiv:1902.01789 [hep-ph]].


\bibitem{Chao:2020qpe}
W.~Chao, Y.~Hu, S.~Jiang and M.~Jin,
[arXiv:2009.14703 [hep-ph]].



\bibitem{pdg}
P.A. Zyla et al. (Particle Data Group), Prog. Theor. Exp. Phys. 2020, 083C01 (2020) and 2021 update.

\bibitem{Esteban:2018azc}
I.~Esteban, M.~C.~Gonzalez-Garcia, A.~Hernandez-Cabezudo, M.~Maltoni and T.~Schwetz,
JHEP \textbf{01}, 106 (2019)
[arXiv:1811.05487 [hep-ph]].

\bibitem{luo2020:neutrino}
S.~Luo,
Phys. Rev. D \textbf{101} (2020) no.3, 033005
[arXiv:1911.06301 [hep-ph]].



\bibitem{Denton:2019pka}
P.~B.~Denton, S.~J.~Parke, T.~Tao and X.~Zhang,
Bull. Am. Math. Soc. \textbf{59}, no.1, 31-58 (2022)
[arXiv:1908.03795 [math.RA]].

\bibitem{Cline:2017qqu}
J.~M.~Cline and J.~M.~Cornell,
Phys. Lett. B \textbf{782}, 232-237 (2018)
[arXiv:1711.10770 [hep-ph]].

\bibitem{Fryer:2006wy}
C.~L.~Fryer, S.~Liu, G.~Rockefeller, A.~Hungerford and G.~Belanger,
Astrophys. J. \textbf{659}, 389-406 (2007)
[arXiv:astro-ph/0609483 [astro-ph]].

\bibitem{Herter:1984}
T.~Herter, J.~R.~Houck, M.~Shure, G.~E.~Gull and P.~Graf
The Astrophysical Journal, 287:L15-L18, 1984

\bibitem{Zurek:2013wia}
K.~M.~Zurek,
Phys. Rept. \textbf{537}, 91-121 (2014)
[arXiv:1308.0338 [hep-ph]].

\bibitem{XENON:2018voc}
E.~Aprile \textit{et al.} [XENON],
Phys. Rev. Lett. \textbf{121} (2018) no.11, 111302
[arXiv:1805.12562 [astro-ph.CO]].

\bibitem{DarkSide-20k:2017zyg}
C.~E.~Aalseth \textit{et al.} [DarkSide-20k],
Eur. Phys. J. Plus \textbf{133} (2018), 131
[arXiv:1707.08145 [physics.ins-det]].

\bibitem{Song:2020nfh}
N.~Song, S.~W.~Li, C.~A.~Arg\"uelles, M.~Bustamante and A.~C.~Vincent,
JCAP \textbf{04}, 054 (2021)
[arXiv:2012.12893 [hep-ph]].

\bibitem{deSalas:2016svi}
P.~F.~de Salas, R.~A.~Lineros and M.~T\'ortola,
Phys. Rev. D \textbf{94} (2016) no.12, 123001
[arXiv:1601.05798 [astro-ph.HE]].

\end{thebibliography}
\end{document}